\documentclass[10pt,journal]{IEEEtran}
\usepackage{mathtools}
\usepackage{amsmath}
\usepackage{amsthm}
\usepackage{amsfonts}
\usepackage{amssymb}
\theoremstyle{definition}

\usepackage{graphicx}
\usepackage{array}
\usepackage{algorithm}
\usepackage{algorithmic}
\usepackage[usenames, dvipsnames]{color}

\newcommand{\eg}{\textit{e.g.},}
\newcommand{\ie}{\textit{i.e.},}

\newcommand{\eqend}{\,.}

\newcommand{\pr}[1]{\mathbb{P}\left\{ #1 \right\}}

\newcommand{\mc}[1]{\mathcal{#1}}
\newcommand{\set}[1]{\left\{ #1 \right\}}
\newcommand{\argmin}{\operatornamewithlimits{argmin}}

\begin{document}
\title{Energy-Efficient Routing in Wireless Networks in the Presence of Jamming}
\author{
    \IEEEauthorblockN{Azadeh Sheikholeslami\IEEEauthorrefmark{1}, Majid Ghaderi\IEEEauthorrefmark{2},\IEEEmembership{  Member, IEEE}, Hossein Pishro-Nik\IEEEauthorrefmark{1},\IEEEmembership{  Member, IEEE},\\ Dennis Goeckel\IEEEauthorrefmark{1}},\IEEEmembership{ Fellow, IEEE}\\
    \IEEEauthorblockA{\IEEEauthorrefmark{1}Electrical and Computer Engineering  Department, University of Massachusetts, Amherst, MA
    \\\{sheikholesla,pishro,goeckel\}@ecs.umass.edu}\\
    \IEEEauthorblockA{\IEEEauthorrefmark{2}Department of Computer Science, University of Calgary
    \\mghaderi@ucalgary.ca}
    \thanks{This work was supported by the National Science Foundation under grants
   CNS-1018464, CNS-0905349 and CIF-1249275. A preliminary version of this work appeared in IEEE ICC 2014 \cite{icc2014}.}
}
\date{}
\maketitle
\begin{abstract}

The effectiveness and simple implementation of physical layer jammers make them an essential threat for wireless networks. In a multihop wireless network, where jammers can interfere with the transmission of user messages at  intermediate nodes along the path, one can employ jamming oblivious routing and then employ physical-layer techniques (e.g. spread spectrum) to suppress jamming. 
However, whereas these approaches can provide significant gains, the residual jamming can still severely limit system performance. 
This motivates the consideration of routing approaches that account for the differences in the jamming environment between different paths.  
First, we take a straightforward approach where an equal outage probability is allocated to each link along a path and develop a minimum energy routing solution. 
Next, we demonstrate the shortcomings of this approach and then consider the joint problem of  outage allocation and routing by employing an approximation to the link outage probability.  This yields an efficient and effective routing algorithm that only requires knowledge of the measured jamming at each node. 
Numerical results demonstrate  that the amount of energy saved by the proposed methods with respect to a standard minimum energy routing algorithm, especially for parameters appropriate for terrestrial wireless networks, is substantial.
\end{abstract}
\section{Introduction}\label{sec:intro}
Due to their broadcast nature, wireless networks  are susceptible to many security attacks. Among them, denial-of-service (DoS) attacks can severely disrupt network performance, and thus are of  interest here. In particular, jamming the physical layer is one of the simplest and most effective attacks, as any cheap radio device can broadcast electromagnetic radiation to block the communication channel \cite{pelechrinis2011denial}.

A straightforward approach to combat adversaries that jam transmissions in the network, particularly in a system with transmitters and receivers capable of operating over a large bandwidth, is to employ physical-layer mitigation techniques. Prominent among these approaches are direct-sequence and frequency-hopped spread spectrum, each of which employs a significantly larger bandwidth than that required for message transmission in order to allow for interference suppression \cite{proakis2000digital,peterson1995introduction}.  
These techniques allow a significant reduction in the impact of the interference, often on the order of the ratio of the system bandwidth to the data rate.  However, the interference can still limit the performance of the system, or, stated differently, spread-spectrum might simply increase the cost of the jamming for the adversary, whom may still be willing to pay such a cost. In addition, the jammers may use alternate methods of jamming to   greatly impact receiver operation by compressing the dynamic range
of the receiver’s front-end \cite{sheikholeslami2014everlasting}.

This motivates the consideration of routing approaches to avoid adversarial jammers  if it can be justified from the perspective of minimizing total cost to the network. In this work, we consider wireless communication between a source and a destination in a multi-hop fashion in the presence of multiple physical layer jammers that are spread over the network area at arbitrary locations by the adversary. 
 We define that cost to be the aggregate energy expended by the system nodes to reliably transmit a message from the source to the destination, with reliability measured by an outage constraint.
The general routing problem has been studied extensively in the literature
 \cite{broch1998performance,royer1999review,draves2004routing,perkins2008ad}. 
Specifically, in \cite{altman2009jamming} and \cite{tague2011jamming}, routing algorithms in the presence of multiple jammers are investigated, but the energy consumption of the network nodes is not considered.
 Excessive energy consumption quickly depletes  battery-powered nodes, and causes increased interference, resulting in a lower network throughput; thus, it is essential to seek  methods to  reduce  energy consumption of the network nodes \cite{feeney2001investigating}. 
There has been some study of energy-aware  routing protocols in the literature \cite{singh1998power,rodoplu1999minimum,chang2000energy,kwon2006energy,dehghan2011minimum}, but only a few works  considered minimum energy routing with security considerations \cite{ghaderi2013efficient,ghaderi2014min}. 
These works studied energy-aware routing in the presence of passive eavesdroppers; however, minimum energy routing in the presence of active adversaries (i.e. jammers) has not been considered.

In this paper, we formulate the minimum energy routing problem with an end-to-end outage probability constraint in a wireless multi-hop network with malicious jammers.  For exposition purposes and the simulation environment, the jammers are assumed to be equipped with omni-directional antennas and to be able to propagate radio signals over the entire frequency band utilized by the nodes in the network.  However, it will become apparent that the proposed algorithms apply in a more general environment, relying only on the measured jamming at each of the nodes in the network and being agnostic of the manner in which that jamming was generated and the geographical locations of the jammers (i.e. the solution easily addresses jammers with directional antennas, etc.).  We will consider both static jammers, which transmit the jamming signal continuously, and simple dynamic jammers that switch randomly between transmitting the jamming signal and sleeping mode. 
  
A difficulty in solving this problem is deciding the local outage of
the links that form a path from source to destination so that the
path satisfies an end-to-end outage requirement. 
We begin our exploration of the multi-hop minimum energy routing problem in the presence of malicious jammers by considering a straightforward approach that allocates equal outage probability to each link along each potential path from source to destination, in such a way that the resulting end-to-end outage probability satisfies a pre-specified threshold.  In this scenario, the search for the optimal path is complicated by a lack of knowledge of the number of hops in the optimal path a priori.  After developing an algorithm to find the optimal path under this approach, we then analyze the potential weaknesses of the solution.  
In particular, if certain links along a path are subject to significant jamming relative to other links along that path, it may be more energy efficient to allow larger outage probabilities on those links subject to significant jamming.  
This motivates a more general approach to the problem where the end-to-end outage constraint is allocated optimally to the links along each path during the process of path selection.  

Unfortunately, the presence of jammers in combination with the end-to-end outage probability constraint makes it difficult to find an optimal path with minimum energy cost.  
{The solution we propose here is to approximate the outage probability
with a simpler expression that allows us to derive an analytical solution for the problem. In fact, the
specific structure of the link cost has a profound impact on the complexity of the routing problem. While the approximate link cost employed in the problem considered here results in a cost structure that is amenable to a polynomial
algorithm, there is no guarantee that, even if the exact link cost had an analytical solution, it would lead
to a polynomial time algorithm (for example, the exact link cost resulted in an NP-hard routing problem
in \cite{ghaderi2014min}). Our simulation results indicate that the gap between the exact and approximate solutions of the routing problem is small.}
In particular, we are able to readily derive a fast and efficient algorithm that, importantly, does not rely on the detailed jammer characteristics (locations, jamming powers) but rather only the observed (and thus measurable) long-term average aggregate interference at each system node. 
 Numerical results are then presented to compare in detail the performance of the various algorithms in terms of energy expended for a given network simulation scenario and end-to-end outage constraint for both single-flow and multiple-flow scenarios. Finally, we discuss how   the proposed algorithm can be implemented in a distributed manner.
 
The rest of the paper  is organized as follows. 
Section \ref{sec:system} describes the system
model. 
The algorithm for minimum energy routing with equal outage per link  is considered in Section \ref{sec:equal}.
The  minimum energy routing with approximate outage per link   in the presence of static and dynamic jammers is presented in Section \ref{sec:proposed}. In Section \ref{sec:numerical}, the results of numerical examples
for various realizations of the system are provided, and the comparison of the
proposed methods to a benchmark shortest path algorithm  is presented.
{Distributed implementation of the routing algorithm and retransmission-aware algorithms are discussed in Section \ref{sec:disc}}, and   
conclusions and ideas for future work are
discussed in Section \ref{sec:conc}.
\section{System Model}\label{sec:system}
\subsection{System Model}
We consider a  wireless network where the system nodes are located arbitrarily.
Let $G=(\mc{N}, \mc{L})$ denote the graph of the network where $\mc{N}$ denotes the set of network nodes and $\mc{L}$ denotes the set of links between them (a link can be potentially formed between any pair of nodes in the network).
 In addition, malicious jammers are  present in the network at arbitrary locations, and these jammers try to interfere with the transmission of the system nodes by transmitting  random signals. 
 We assume that each jammer utilizes an omni-directional antenna and  can transmit over the entire frequency band;  thus, spread spectrum or frequency hopping strategies  improve performance via the processing gain, but are not completely effective in interference suppression. 
 
 One of the system nodes (source) chooses  relays,  with which it conveys its message to the destination in a (possibly) multi-hop fashion.
Suppose the relays that the source selects construct a $K$-hop route between the source and the destination.
A $K$-hop route $\Pi$ is determined by a set of $K$ links $\Pi=\left\langle \ell_1,\ldots,\ell_K\right\rangle$ and $K+1$ nodes (including source and destination) such that link $\ell_k$ connects the $k^{th}$ link transmitter $S_k$ to the $k^{th}$ link receiver $D_k$.

In this work we consider a delay-intolerant network, which is a common assumption especially in military networks.
	If we enable retransmissions at relays, the local retransmissions cause  out of control returns of the message between relays, and thus  impose undesirable delay on the network.
	Hence, we do not consider local retransmissions in this paper.
 
We denote the set of jammers  by $\mathcal{J}$ and consider both static jammers and dynamic jammers. 
In the case of static jammers,  each jammer transmits white Gaussian noise with a fixed power.  
Since the jammers are active, we assume initially that the transmit power and the location of jammers are known to the system nodes; 
however, we will see that for our proposed method, \textit{the knowledge of the transmit powers and locations of jammers is not necessary};
 in fact, the system nodes can measure the average received jamming (averaged over the multipath fading)  and use this estimate of jamming interference for efficient routing. 
In the case of dynamic jammers, each jammer switches between an ``ON'' state, when it transmits the jamming signal,  and an ``OFF'' state or sleeping mode randomly and independently from the other jammers. 
These dynamic jammers are especially useful when the battery life of the jammers is limited and the adversary tries to cover a larger area, as the jammers in  sleep mode can save significant energy.
\subsection{Channel Model}
We assume frequency non-selective Rayleigh fading between any  pair of nodes. 
 For instance, for link $k$ between nodes $S_k$ and $D_k$, let $h_k$ denote the  fading,  and $\{h_{j,k}\}_{j\in\mathcal{J}}$ denote the respective fading coefficients  between jammers and  $D_k$. 
It follows that the channel fading power is exponentially distributed. Without loss of generality, we assume  $E[|h_{k}|^2]=1,\; \forall k$, and $E[|h_{j,k}|^2]=1,\; \forall j,k$, and then work path-loss explicitly into (\ref{eq:1}) below.
Also, each receiver experiences additive white Gaussian noise with power $N_0$. Hence, the signal received by node $D_k$ from node $S_k$ is
\begin{equation}\label{eq:1}
y^{(k)}=\frac{h_{k}\sqrt{P_k}}{d_{k}^{\alpha/2}}x^{(k)}+\sum_{j\in \mathcal{J}}\frac{h_{j,k}\sqrt{P_j}}{d_{j,k}^{\alpha/2}}x^{(j)}+n^{(k)},
\end{equation}
where  $P_k$ is the transmit power of node $S_k$, $P_j$ is the transmit power of the $j^{th}$ jammer,  $d_{k}$ is the distance between  $S_k$ and $D_k$, $d_{j,k}$ is the distance between  $j$-th jammer and $D_k$, and $\alpha$ is the path-loss exponent.
Also, $x^{(k)}$ and $x^{(j)}$ are the unit power signals transmitted by $S_k$ and $j$-th jammer.
If spread spectrum were employed, the model would obviously change to include the processing gain and further averaging of the fading, but the design process would be similar. 
\begin{small}
\renewcommand{\arraystretch}{1.5}
\begin{table}
	\begin{center}
	\caption{Table of notations}\label{tab:1}
		\begin{tabular}[c]{|l| p{7.2cm}|}
		\hline
		$\pi$& Desired end-to-end outage probability   \\
		\hline
		$p^{SD}_{out}$&The average source-destination (i.e., end-to-end) outage probability\\
		\hline
		$p^k_{out}$&The average outage probability of  $k^{th}$ link\\
		\hline
		$h_k$ & Fading coefficient of  $k^{th}$ link   \\
		\hline
		$h_{j,k}$ & Fading coefficient between $j^{th}$ jammer and   the receiver node of link $k$   \\
		\hline
		$d_k$ & The distance between the transmitter and receiver of link $k$  \\
		\hline
		$d_{j,k}$ & The distance between $j^{th}$ jammer and  $k^{th}$ receiver node   \\
		\hline
		$P_k$& Transmit power of  the  transmitter on $k^{th}$ link \\
		\hline
			$P_j$&Transmit power of $j^{th}$ jammer\\
			\hline
			$J_k$&Expected value of the total received power at the receiver of link $k$ from jammers\\
			\hline
			$\mathcal{J}$&Set of jammers in the network\\
			\hline
			$\mc{N}$&Set of network nodes\\
			\hline
			$N$&Number of network nodes\\
			\hline
			$\mc{L}$&Set of  links of the network \\
			\hline
			$\mc{C}(.)$&Cost of establishing the argument (link or path)\\
			\hline
		$\alpha$&Path-loss exponent  \\
		\hline
			$\gamma$& The required signal-to-interference ratio at each receiver\\
			\hline
			$N_0$&Thermal noise power\\
			\hline
		\end{tabular}
\end{center}
	\end{table}
	\end{small}
\subsection{Path Outage Probability}
Our goal is to find a minimum energy route  between  an arbitrary pair of nodes in the network  such that the desired average end-to-end  probability of outage is guaranteed.  
Hence, we need to find the set of relay nodes (links) with minimum aggregate power such that the end-to-end probability of outage $p_{out}^{SD}\leq \pi$, where $\pi$ is a predetermined threshold for the average outage  probability.
Let $p_{out}^{k}$ denote the average outage probability of link $\ell_k=\left\langle S_k,D_k\right\rangle$; the source-destination outage probability in terms of the outage probability of each link is,
\begin{align} \label{eq:pout1}
p_{out}^{SD}&=  1-\prod_{1\leq k\leq K}\left(1-p_{out}^{k} \right).
\end{align}
Implicit in our formulation is the end-to-end throughput of the path between the source and destination. Let $\rho$ denote the required end-to-end throughput. Since the throughput of a path is determined by the throughput of its bottleneck link, to minimize transmission energy of the path, it is necessary to achieve an equal throughput over each link of the path. Thus, in our formulation of minimum energy routing, the cost of each link is computed with respect to the required throughput $\rho$, as described in the following subsection.   
\subsection{Analysis of Link Outage Probability}\label{sec:outageprobability}
Consider the outage probability of a link in the presence of the set of  jammers $\mc{J}$. 
The outage probability of link $\ell_k$ given its fading gain $|h_{k}|^2$ and the fading gains  between the jammers and the receiver of the link, i.e.,  $\{|h_{j,k}|^2\}_{j\in\mathcal{J}}$ is,
\begin{align} \label{eq:outage}
p_{out}^{k}&=\pr{\frac{P_{k}|h_{k}|^2/d_{k}^\alpha}{N_0+\sum_{j\in \mathcal{J}} P_{j}|h_{j,k}|^2/d_{j,k}^\alpha}<\gamma },
\end{align}
where  $\gamma$ is the required signal-to-interference ratio at the receiver. The value of $\gamma$ determines the link throughput. Specifically, for a desired throughput of $\rho$, by applying the Shannon capacity formula, the threshold $\gamma$ is given by:
\[
 \gamma = 2^\rho - 1.\]
Since the fading gain $|h_{k}|^2$ is distributed exponentially, conditioned on $\{|h_{j,k}|^2\}_{j\in\mathcal{J}}$, we obtain that, 
\begin{small}
\begin{equation}
\label{eq:con_outage}
p^k_{out}(\{|h_{j,k}|^2\}_{j\in\mathcal{J}})=  1-\exp\left(\frac{-\gamma\left(N_0+\sum_{j\in \mathcal{J}} P_{j}|h_{j,k}|^2/d_{j,k}^\alpha\right)}{P_{k}/d_{k}^\alpha}\right).
\end{equation}
\end{small}
Taking the expectation over the fading gains of the jammers yields:
\begin{small}
\begin{align}\nonumber 
p_{out}^{k}&= E\left[1-\exp\left(\frac{-\gamma\left(N_0+\sum_{j\in \mathcal{J}} P_{j}|h_{j,k}|^2/d_{j,k}^\alpha\right)}{P_{k}/d_{k}^\alpha}\right)\right]\\\nonumber
&=1-e^{-\frac{\gamma N_0d_{k}^\alpha}{P_{k}}}\prod_{j\in \mathcal{J}}E\left[\exp\left(\frac{-\gamma P_{j}|h_{j,k}|^2/d_{j,k}^\alpha}{P_{k}/d_{k}^\alpha}\right)\right]\\\label{eq:exact}
&=1-\frac{e^{-\frac{\gamma N_0d_{k}^\alpha}{P_{k}}}}{\prod_{j\in \mathcal{J}}\left(1+\frac{\gamma P_{j}/d_{j,k}^\alpha}{P_{k}/d_{k}^\alpha}\right)},
\end{align}
\end{small}
which is the expected outage probability for a link in the network. 
The last equality follows from the fact that if the random variable $X$  is exponentially distributed, $E[e^{-tX}]=\frac{1}{1+t\lambda^{-1}}$ where  $\lambda=E[X]$ and $t\in \mathbb{R}$.
 \subsection{Minimum Energy Routing: the Optimization Problem}
Our goal is to find the  path that connects the source and destination  with minimum energy consumption for the communication subject to an
end-to-end outage probability constraint. 
The minimum energy routing problem is to find the optimal path $\Pi^*$ so that:
\begin{equation}\label{eq:route}
\Pi^* = \argmin_{\Pi \in {\bold{\Pi}_{SD}}} \mc{C}(\Pi)
\end{equation}
where, $\bold{\Pi}_{SD}$ denotes the set of all possible paths between source and destination nodes $S$ and $D$, and $\mc{C}(\Pi)$ is the minimum cost to establish path $\Pi$, which is given by the following optimization problem:
\begin{equation}\label{eq:cost1}
\mc{C}(\Pi)=\min_{k=1,\underset{ P_k > 0}\ldots,K,}\sum_{\ell_k\in\Pi} P_k, \quad\text{s.t.,}\quad p_{out}^{SD}(\Pi)\leq \pi.
\end{equation}
$\ell_k$ refers to link $k$ in path $\Pi = \langle \ell_1, \ell_2, \ldots, \ell_{|\Pi|}\rangle$ (where, $|\Pi|$ denotes the length of path $\Pi$). Substituting (\ref{eq:exact})  in (\ref{eq:pout1}), the constraint of this optimization problem is,
\begin{equation}\label{eq:pout22}
p_{out}^{SD}(\Pi)=1-\prod_{\ell_k\in\Pi}\frac{e^{-\frac{\gamma N_0d_{k}^\alpha}{P_{k}}}}{\prod_{j\in \mathcal{J}}\left(1+\frac{\gamma P_{j}/d_{j,k}^\alpha}{P_{k}/d_{k}^\alpha}\right)}\leq \pi.
\end{equation}
{Notice that classical routing algorithms such as Dijkstra or Bellman-Ford cannot be applied to this problem  as they require an explicit characterization of the cost of each link, which is not possible in this problem as the cost of a link actually depends on the path that contains the link. 
That is, the link costs depend on the unknown path between $S$ and $D$. In order to determine the cost of a link, first we have to determine the path that contains the link. 
Consider some link $\ell$ that is on two possible paths $\Pi_1$ and $\Pi_2$. 
The paths $\Pi_1$ and $\Pi_2$ have at least one link that is not on both of them, otherwise they are just the same path. 
The characteristics of the uncommon links change the distribution of the outage probability among the links of each path. 
Thus, the cost of link $\ell$, which is the power allocated to its source node, depends on the end-to-end path that contains link $\ell$. Depending on which path is considered, the cost of the link changes. Such a structure is completely different from the structure required for classical routing algorithms to be applied in a network.
In general, there are two approaches to solve this problem:
\begin{enumerate}
	\item Exploit the structure of the problem in order to design a solution that is efficient (\eg\ has polynomial running time).
	\item Ignore the problem structure and solve it numerically, which may or may not be efficient depending on the structure of the problem.
\end{enumerate}}

{While the second approach will work and could be used to find the optimal path, we wish to comment on the computational complexity of this approach. To numerically solve this problem, one has to find all possible paths between nodes $S$ and $D$, and then choose the one that has the lowest cost and satisfies the outage constraint. 
Finding all possible paths, or for that matter even counting their number, between a pair of nodes in a network is a well-known combinatorial problem with no known polynomial solution. It belongs to the class of problems known as \mbox{\#P-complete}, and is even hard to approximate \cite{valiant1979complexity}. 
So, while it is possible to numerically solve problem defined in (\ref{eq:route}), (\ref{eq:cost1}), and (\ref{eq:pout22}) for a small network, the running time of such an approach will be prohibitive for any large network of interest.}

{It may be  possible to design a psudo-polynomial time (\ie\ exponential time in the length of the input) algorithm to solve problem (\ref{eq:route})-(\ref{eq:pout22}) exactly. To this end, we note the similarity between our problem and the delay-constrained routing problem defined as follows:}

\textbf{{Delay-Constrained Routing Problem:} }

{\begin{equation}
\label{e:delay}
\begin{split}
\Pi^* = &\arg\min_{\Pi \in \bold{\Pi}_{SD}} \sum_{\ell_k \in \Pi} C_k\\
&\text{Subject to:}\\
& \qquad \pr{\text{end-to-end delay on path $\Pi$}} \le \tau
\end{split}
\end{equation}}
{where, $C_k$ is the cost of link $\ell_k$. It has been shown that the Delay-Constrained Routing problem is NP-complete \cite{wang1996quality}. }

{In the paper, we will take the following approaches in order to design an algorithm with polynomial time complexity.}
\begin{enumerate}
	\item 
	As a reasonable algorithm to help motivate our main approach, we first simplify the problem and consider equal outage probabilities per-link    such that the end-to-end outage probability over the path is $\pi$,  which is described in the next section. However, we show that  this approach could lead to severe inefficiencies.  
\item
Thus, we use an approximation to tackle the complexity of the optimization problem defined in (\ref{eq:route}), (\ref{eq:cost1}), and (\ref{eq:pout22}). Using the approximation, we develop an algorithm to find the efficient route. 
\end{enumerate}

\section{MER-EQ: Minimum Energy Routing with Equal Outage per Link}\label{sec:equal}

 
As explained earlier, in this approach,  we simplify the problem and consider equal outage probabilities per links of the optimum path such that the desired end-to-end outage probability $\pi$ is guaranteed.
If the optimum path has $h$ hops, assuming equal outage per link, the per-hop outage probability is,
\begin{equation}\label{eq:eps}
    \varepsilon(h) = 1 - \sqrt[h]{1-\pi}
    \eqend
\end{equation}

 Let $\mc{C}({u,v})$ denote the cost of the link between nodes $u$ and $v$. 
The cost of establishing one link is a function of the outage probability of that link, which in turn is dependent on the path length $h$. We use the notation $P_{u,v}(\varepsilon(h))$ to denote the transmission power required for link $\ell_{u,v}$ when the link is part of a path of length $h$.
However, a difficulty of this approach is that the number of links of the optimum path  is not  known a priori, and thus the per link outage probability $\varepsilon(h)$ is not known.
This means that, in order to compute the cost of each link, we need to have the optimal path, but in order to find the optimal path, we need to compute the cost of each link. 
Because of the interdependency of link costs and optimal path, traditional routing algorithms such as Dijkstra's algorithm cannot be applied to this problem. 
We need to design an algorithm where the cost of a link depends on the length of the path. 

To this end, we develop a two-step algorithm as follows. 
In the first step, we assume the number of hops is $h$, and then we calculate the per-hop outage probability by applying (\ref{eq:eps}).  Using this per-hop outage probability, we calculate the cost of establishing each link assuming the link is on a path of length $h$ from source to destination.  However, even with these link costs calculated, it is not trivial to perform shortest path routing under the constraint that the route found must have $h$ hops, since standard shortest path algorithms (such as Dijkstra) do not enforce such a constraint.  Hence, we do the network expansion described in the next section before running a standard shortest path algorithm to complete the first step of MER-EQ.  We repeat the first step for each possible number of hops $h=1,2,\ldots,N-1$.  The second step produces the output of MER-EQ by selecting the route with minimum energy among the $N-1$ paths, one for each $h=1,2,\ldots,N-1$, obtained in the first step.

\subsection{Selection of a Minimum Cost Path of Length $h$ Hops }
\begin{figure}
\begin{center}
	\includegraphics[width=.48\textwidth]{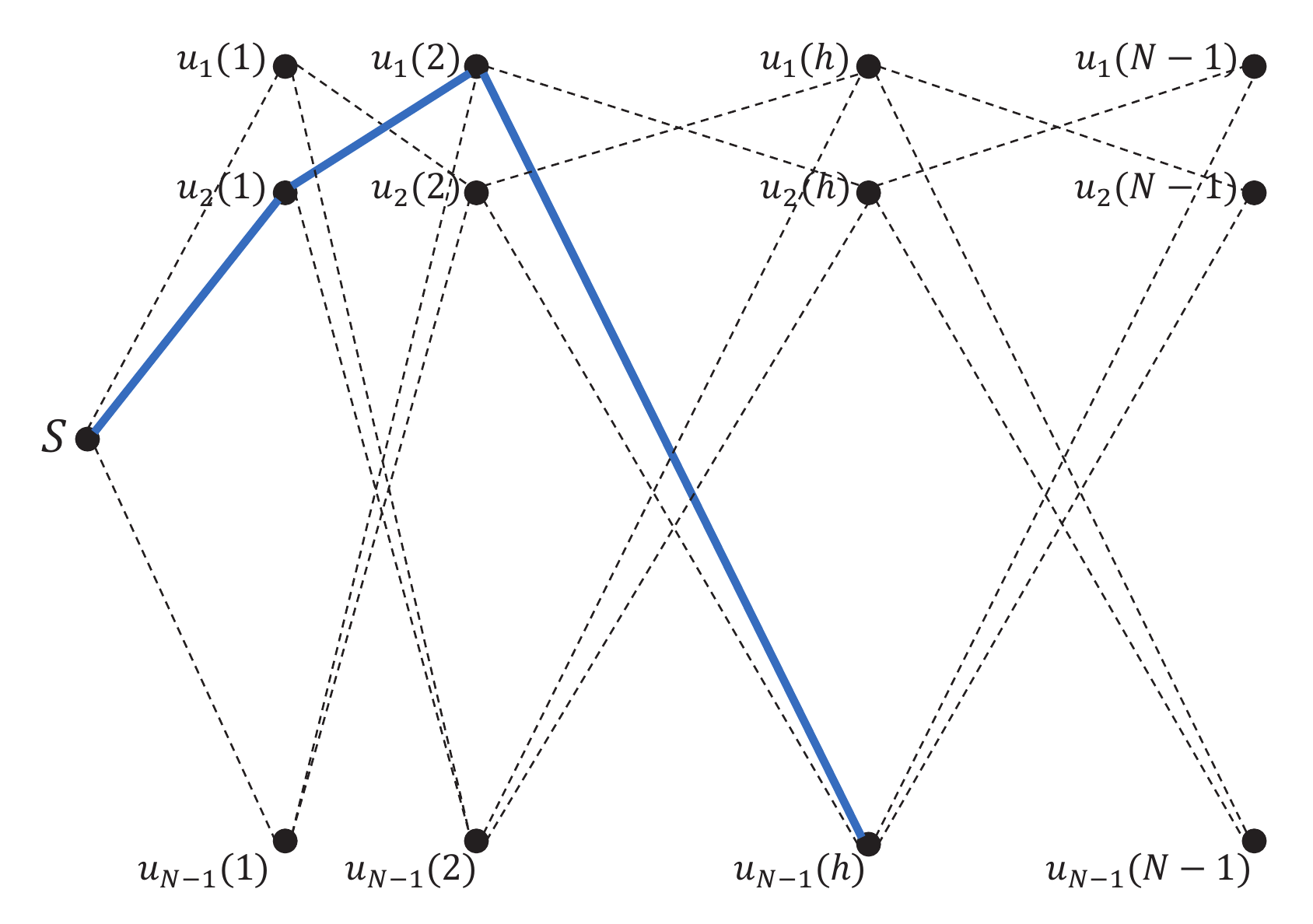}
 \end{center}
 \caption{Network expansion:   add $N-1$ replicas for each node $u_i,\: i=1,\ldots,N-1$ to the expanded network. Then links (shown by dashed lines) are added to the expanded network such that a path from $S$ to $u_i(h)$ will have exactly $h$ hops. Hence, every path from $S$ to $D(h)$ has exactly $h$ hops. A sample path from the source to $u_{N-1}(h)$ is shown by bold solid lines. }
 \label{fig:network_expansion}
 \end{figure}

To enforce the selection of a route with $h$ hops as required in the first step of MER-EQ, we pre-process the network to create an expanded network as described in Algorithm~\ref{a:expansion}. In this algorithm, $S$ and $D$ denote the source and destination nodes. The algorithm works by first adding $S$ to the expanded network. Next, since the longest path in a network of $N$ nodes will have at most $N-1$ hops,  it adds $N-1$ replicas for each node $u_i,\: i=1,\ldots,N-1$ to the expanded network.
Let us denote the $h^{th}$ replica of node $u_i$ by $u_i(h)$.  Then links are added to the expanded network such that a path from $S$ to $u_i(h)$ will have exactly $h$ hops (Figure \ref{fig:network_expansion}). 
Similarly, every path from source $S$ to $D(h)$ has $h$ hops.
Consequently, the shortest path from $S$ to $D(h)$ in the expanded network has  precisely $h$ hops.

\begin{algorithm}[t]
\caption{Network Expansion($G=(\mc{N}, \mc{L})$)}
\renewcommand{\algorithmiccomment}[1]{/*~#1~*/}
\begin{algorithmic}[1]
    \STATE $\mc{N}' = \set{S}$
    \STATE $\mathcal{L}'=\{\}$
    \STATE\COMMENT{replicate every node of the original graph to $N-1$ nodes (except source)}
    \FORALL{$u \neq s \in \mc{N}$}
        \STATE $\mc{N}' = \mc{N}' + \set{u(1), u(2), \ldots, u(N-1)}$
    \ENDFOR
     
    \STATE\COMMENT{connect source node $S$ to every $u(1)$ node}
    \FORALL{$\ell_{S,u} \in \mc{L}$}
        \STATE $\mc{L}' = \mc{L}' + \ell_{S,u(1)}$
    \ENDFOR
    \STATE\COMMENT{connect every $u(h)$ to every $v(h+1)$ node ($u\neq v$)}
    \FORALL{$\underset{u\neq s, u\neq d, v\neq s}{\ell_{u,v}} \in \mc{L}$}
        \FOR{$h=1$ \TO $N-2$}
            \STATE $\mc{L}' = \mc{L}' + \ell_{u(h), v(h+1)}$
        \ENDFOR
    \ENDFOR
    \RETURN $G'=(\mc{N}', \mc{L}')$
\end{algorithmic}
\label{a:expansion}
\end{algorithm}


\subsection{Routing Algorithm}
The routing algorithm is described in Algorithm \ref{a:routing}. To compute  the minimum cost path, first we find the shortest path for every number of hops, $h = 1, \ldots, N-1$  in the expanded network by repeatedly employing   Dijkstra's algorithm (line 7). 
Then, the algorithm chooses the path with minimum cost from source to destination and returns the optimum path and its cost (lines 11 and 12).
This path is computed by finding the least cost path among the paths that have $h = 1, 2, .., N-1$ hops. Let $\Pi(h)$ denote the minimum cost path of length h between the source and destination. Then, the optimal path is computed as follows:
\[
\Pi^* = \arg \min_{h} C(\Pi(h)).
\]
\vspace{-17pt}
\subsection{Discussion} \label{sec:disc}
\begin{figure}
\begin{center}
	\includegraphics[width=.35\textwidth]{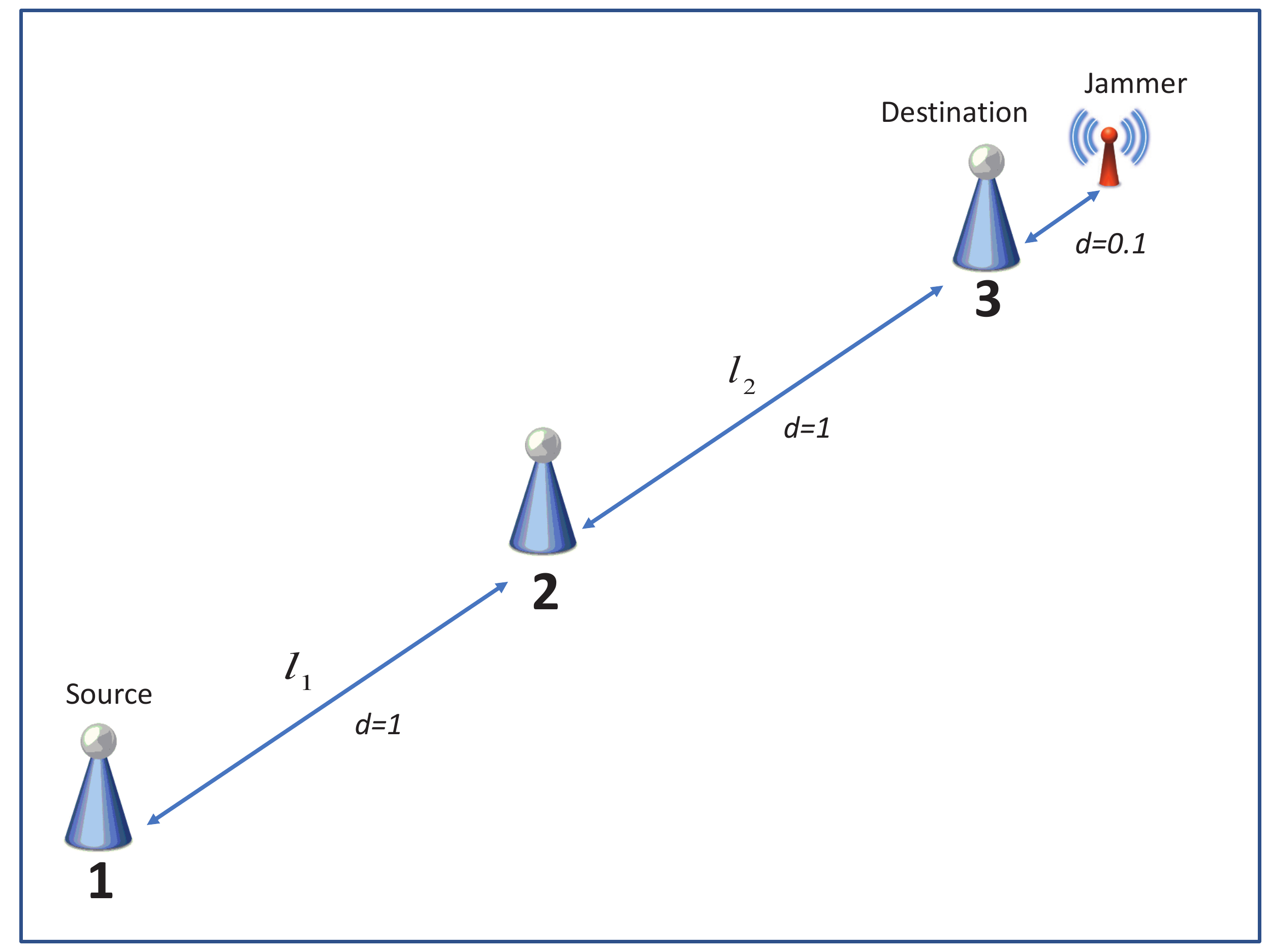}
 \end{center}
 \caption{A wireless network in the presence of one jammer is shown here. In this network, by allocating unequal outage probability to different links, the cost  of communication decreases significantly. }
 \label{fig:toy}
 \end{figure}

The algorithm described in this section is not efficient, since we force all links to have the same outage probability. This limitation can increase the cost of communication unnecessarily.
For example, consider a network in the presence of one  jammer in Fig. \ref{fig:toy}.
Suppose that the end-to-end outage probability $p_{out}^{SD}=0.1$,  path-loss exponent $\alpha=2$,  jamming power $P_j=1$, $N_0=1$, and $\gamma=1$.
By using the MER-EQ routing algorithm, the minimum-energy path from the source to the destination is a two-hop path. 
In this case, in order to obtain $p_{out}=0.1$, the outage probability of each link  $p_{out}^1=p_{out}^2=0.051$. 
Hence, from (\ref{eq:exact}) the transmit  power of the source node is $P_1=34.5$, and the transmit power of node 2 is $P_2=1868.2$, and thus the total power is $P=1902.7$. 
The reason that $P_2$ is  so high is the interference from the near jammer.
However, if we change the outage probability allocation between the two links, and allow the transmission between node 2 and the destination to have a larger outage probability, we expect that the aggregate power consumption decreases.
For instance, suppose the outage probability of link $\ell_1$ is $p_{out}^1=0.01$ and the outage probability of link $\ell_2$ is $p_{out}^2=0.0909$.  
In this case, from  (\ref{eq:exact}), the transmit power of the source node is $P_1=181.5$ and the transmit power of node 2 is $P_2=1011.1$, and thus the total power is $P=1192.6$.
We see that by relaxing the restriction on the allocation of the outage probability between different links, the cost of communication decreases significantly.

Moreover, in order to find the optimal path we basically need to apply the shortest path algorithm $N-1$ times, which makes this approach inefficient in term of running time in large networks.
Each application of the Dijkstra's algorithm in the expanded network requires a running time of $O(N^2 \log N)$, and thus the algorithm MER-EQ takes $O(N^3 \log N)$ time to run.

In the remainder of the paper, we present a minimum energy routing algorithm with approximate outage per link and demonstrate how using an estimate of the end-to-end outage probability  leads to a fast and efficient algorithm that  improves the energy efficiency of the network significantly.
\begin{algorithm}[t]
\caption{MER-EQ($G'=(\mc{N}', \mc{L}')$)}
\renewcommand{\algorithmiccomment}[1]{/*~#1~*/}
\begin{algorithmic}[1]
    \FOR{$h=1$ \TO $N-1$}
        
        \STATE \COMMENT{for each link, set the link cost to the transmit power required to maintain the outage probability $\varepsilon(h)$ on the link}
        \FORALL{$\ell_{u,v} \in \mc{L}'$}
            \STATE $\mc{C}({u,v}) = P_{u,v}(\varepsilon(h))$
        \ENDFOR
        
        \STATE\COMMENT{compute the shortest $h$-hop path}
        \STATE $[\Pi(h), C(h)] =$ Dijkstra($G', s, d(h)$)
				\STATE\COMMENT{store the path and its cost in $\Pi(h)$ and $C(h)$}
        
    \ENDFOR
    
    \STATE\COMMENT{choose the best path for reaching the destination}
    \STATE $h^* = \displaystyle \arg \min_{h} C(h)$
    \RETURN $[\Pi(h^*), C(h^*)]$
\end{algorithmic}
\label{a:routing}
\end{algorithm}

 \vspace{-8pt}
\section{MER-AP: Minimum Energy Routing with Approximate Outage per Link}\label{sec:proposed}
In this section, we present our minimum energy routing algorithm with approximate outage per link  by considering the end-to-end outage constraint.
From (\ref{eq:exact}), the per-hop outage probability  $p_{out}^{k}$ is,
\vspace{-6pt}
\begin{align}\nonumber 
p_{out}^{k}&=1-\frac{e^{-\frac{\gamma N_0d_{k}^\alpha}{P_{k}}}}{\prod_{j\in \mathcal{J}}\left(1+\frac{\gamma P_{j}/d_{j,k}^\alpha}{P_{k}/d_{k}^\alpha}\right)}\\
\nonumber &\leq 1-\frac{e^{-\frac{\gamma N_0d_{k}^\alpha}{P_{k}}}}{\prod_{j\in \mathcal{J}}e^{\frac{\gamma P_{j}/d_{j,k}^\alpha}{P_{k}/d_{k}^\alpha}}}\\\label{eq:pout}
&= 1-\frac{e^{-\frac{\gamma N_0d_{k}^\alpha}{P_{k}}}}{e^{\sum_{j\in \mathcal{J}}\frac{ \gamma P_{j}/d_{j,k}^\alpha}{P_{k}/d_{k}^\alpha}}},
\end{align}
where the inequality is from the fact that $e^x\geq 1+x$ for  $x\geq 0$.
 While this is a conservative estimate of the end-to-end outage probability, our simulation results show that it results in an effective solution that results in significant energy savings. 
From (\ref{eq:pout}) we have,
\begin{equation}\label{eq:poutapprox}
p_{out}^k\leq 1-e^{-\frac{\gamma d_k^{\alpha}}{P_k}(N_0+J_k)},
\end{equation}
where $J_k$ is the expected value of the total received power at node $D_k$ from all jammers, i.e. $J_k=\sum_{j\in \mathcal{J}} P_j/d_{j,k}^{\alpha}$.
Importantly, this approximation not only  enables the development of an efficient routing algorithm, but  also  simplifies the implementation of the algorithm in real networks. While the exact outage probability as given in (\ref{eq:exact}) requires the knowledge of jammer powers and their locations, the approximation in (\ref{eq:poutapprox}) requires only the knowledge of the ``average'' jamming power received at a node, which can be readily measured.
\subsection{Approximate Cost of a Given Path}\label{sec:optimalcost}

 Our objective is to find the efficient  path  and the minimum transmission power required to
establish the path to satisfy the outage probability $\pi$,
First, we find the  power allocation for a given path $\Pi$, and then use this result to design a routing algorithm to find the  path. 
To this end, the  power allocation problem for a given path $\Pi=\left\langle \ell_1, ..., \ell_K\right\rangle$ is described by the following optimization problem: 
\[
\min_{k=1,\underset{P_k>0}\ldots,K}\sum_{\ell_k\in\Pi} P_k,
\]
subject to:
\[
p_{out}^{SD}=1-\prod_{\ell_k\in\Pi}(1-p_{out}^k)\leq \pi.
\]
From (\ref{eq:poutapprox}) the equivalent constraint is,
\begin{equation}\label{eq:equicons}
\sum_{\ell_k\in\Pi}d_k^{\alpha}\left(\frac{N_0+J_k}{P_k}\right)\leq \epsilon=\frac{-\ln(1-\pi)}{\gamma}.
\end{equation}
Since the left side of (\ref{eq:equicons}) is a decreasing function of $P_k$ and our goal is to find the route with minimum cost, the inequality constraint can be substituted by the following equality constraint,
 \begin{equation}\label{eq:equicons2}
\sum_{\ell_k\in\Pi}d_k^{\alpha}\left(\frac{N_0+J_k}{P_k}\right)= \epsilon.
\end{equation}
To find the  link costs, we use the Lagrange multipliers technique.
Thus, we need to solve (\ref{eq:equicons2}) and the following $K$ equations simultaneously,
\begin{align*}
\frac{\partial }{\partial P_i} \left\{\sum_{\ell_k\in\Pi} P_k+\lambda \left(\sum_{\ell_k\in\Pi}d_k^{\alpha}\left(\frac{N_0+J_k}{P_k}\right) -\epsilon\right)\right\}=0&,\\ i=1,\ldots,K&.
\end{align*}
Taking the derivative, we obtain that,
\begin{align}
1-\lambda d_i^{\alpha}\frac{(N_0+J_i)}{P_i^2}=0,\quad i=1,\ldots,K,
\end{align}
and thus,
\begin{equation}\label{eq:P_k2}
P_i=\sqrt{\lambda d_i^{\alpha}(N_0+J_i)}.
\end{equation}
On substituting $P_i$ from (\ref{eq:P_k2}) into (\ref{eq:equicons2}), we have,
\begin{equation}\label{eq:lambda}
\lambda=\frac{1}{\epsilon ^2}\left(\sum_{\ell_k\in\Pi} \sqrt{d_k^{\alpha}(N_0+J_k)} \right)^2.
\end{equation}
Hence, by substituting $\lambda$ from (\ref{eq:lambda}) into (\ref{eq:P_k2}), the  cost of each link is given by,
\begin{equation}\label{eq:P_k}
P_i=\frac{1}{\epsilon}\sqrt{d_i^{\alpha}(N_0+J_i)}\sum_{\ell_k\in\Pi}\sqrt{d_k^{\alpha}(N_0+J_k)},
\end{equation}
and the  cost of path $\Pi$ is given by,
\begin{equation}\label{eq:pathcost1}
\mathcal{C}(\Pi)=\frac{1}{\epsilon}\left( \sum_{\ell_k\in\Pi}\sqrt{d_k^{\alpha}(N_0+J_k)}\right)^2.
\end{equation}
Note that the cost of establishing each link depends on the summation of noise power and the expected received jamming signal $N_o+J_k$, and thus in order to calculate cost of each link we do not even need to separate the jamming signal from the noise.
\subsection{Routing Algorithm}
The  path cost structure in (\ref{eq:pathcost1}) allows us to find the minimum energy route from source to destination as follows. First assign the link weight $\mathcal{C}(\ell_k)=\sqrt{d_k^{\alpha}(N_0+J_k)}$ to each potential link $\ell_k$ in the network. Now apply any classic shortest-path algorithm such as the Dijkstra's algorithm.  
This path minimizes the end-to-end weight $\sum_{\ell_k\in\Pi}\sqrt{d_k^{\alpha}(N_0+J_k)}$ and thus it will also minimize the source-destination path cost $\mathcal{C}(\Pi)$ in (\ref{eq:pathcost1}). 
We note that the running time of this algorithm, referred to as MER-AP, is in $O(N \log N)$ as it essentially invokes the Dijkstra's algorithm once.

Now, each node in route $\Pi$ transmits the message to the next node until it reaches the destination. The transmit power of each node is determined by (\ref{eq:P_k}) and the actual  outage probability of each link can be obtained from (\ref{eq:poutapprox}). 

\subsection{Heuristic Adjustment of Transmit Powers}
Consider the optimum route $\Pi$ that is found by applying the MER-AP algorithm. Suppose that route $\Pi$  consists of $H$ hops, 
 and   its achieved end-to-end outage probability is $p_{out}^{SD}$.
Since we consider an upper bound for the end-to-end outage probability in developing MER-AP, the achieved end-to-end outage probability $p_{out}^{SD}$ might be less than the allowed outage probability $\pi$, 
\begin{equation}\label{eq:11}
p_{out}^{SD}\leq \pi,
\end{equation}
Consequently, MER-AP  with the $P_i$'s set as in (\ref{eq:P_k}) can be too conservative  in some instances.
In order to address this, we apply the following heuristic.
Let $\delta$ be the ratio of the actual end-to-end success probability  $1-p_{out}^{SD}$ to the desired success probability $1-\pi$. 
From (\ref{eq:11}),
\[\delta =\frac{1-\pi}{1-p_{out}^{SD}}\leq 1.\]
Now suppose that we set  a new  success probability for each link in the efficient route by multiplying the success probability of each link by a factor  $\sqrt[H]{\delta}$. 
Hence, the new success probability of each link in the  route is $\sqrt[H]{\delta}(1-p_{out}^k)$, which is less than the old success probability of that link since $\sqrt[H]{\delta}\leq 1$. 
By using this approach,  we reduce the required success probability of each link, and thus from (\ref{eq:exact}), the cost of establishing each link decreases, which results in less energy consumption of the algorithm MER-AP.
In this case, the new end-to-end success probability can be calculated as,
\begin{align*}
	&\prod_{k=1,\ldots,H}{\sqrt[H]{\delta}(1-p_{out}^k)}\\
	&=\delta\prod_{k=1,\ldots,H}{(1-p_{out}^k)}\\
	&=\delta{(1-p_{out}^{SD})}=1-\pi,
\end{align*}
which is equal to the desired source-destination success probability. 
Hence, by applying this heuristic, the resultant end-to-end outage probability will be equal to the allowed outage probability while the aggregate cost of communication on the path selected by MER-AP will be less than when we do not apply this heuristic.

\subsection{Routing in the Presence of Dynamic Jammers}
 In this section, we consider the case of dynamic jammers, where each jammer alternates between the jamming mode and the sleeping mode. We model the probabilistic behavior of  jammers by  i.i.d. Bernoulli random variables $\beta_j, j\in\mathcal{J}$, such that $p(\beta_j=1)=1-p(\beta_j=0)=q$. 
 Using (\ref{eq:outage}), the average outage probability of  link $\ell_k$ is:
\begin{small}
 \begin{align}\nonumber 
&p_{out}^{k}= E\left[1-\exp\left(\frac{-\gamma\left(N_0+\sum_{j\in \mathcal{J}} P_{j}\beta_j|h_{j,k}|^2/d_{j,k}^\alpha\right)}{P_{k}/d_{k}^\alpha}\right)\right]
\\\nonumber
&=1-e^{-\frac{\gamma N_0d_{k}^\alpha}{P_{k}}}\prod_{j\in \mathcal{J}}E\left[\exp\left(\frac{-\gamma P_{j}\beta_j|h_{j,k}|^2/d_{j,k}^\alpha}{P_{k}/d_{k}^\alpha}\right)\right]\\\nonumber
&=1-e^{-\frac{\gamma N_0d_{k}^\alpha}{P_{k}}}\prod_{j\in \mathcal{J}}
\Bigg\{qE\left[\exp\left(\frac{-\gamma P_{j}|h_{j,k}|^2/d_{j,k}^\alpha}{P_{k}/d_{k}^\alpha}\right)\right]+(1-q)\Bigg\}\\\nonumber
&=1-e^{-\frac{\gamma N_0d_{k}^\alpha}{P_{k}}}\prod_{j\in \mathcal{J}}
\left\{\frac{q}{\left(1+\frac{\gamma P_{j}/d_{j,k}^\alpha}{P_{k}/d_{k}^\alpha}\right)}+1-q\right\}\\\label{eq:pout2}
&\leq 1-e^{-\frac{\gamma N_0d_{k}^\alpha}{P_{k}}}\prod_{j\in \mathcal{J}}e^{-\frac{\gamma qP_{j}/d_{j,k}^\alpha}{P_{k}/d_{k}^\alpha}},
\end{align}
\end{small}
where the expectations are computed over $\{\beta_j\}_{j\in\mathcal{J}}$ and $\{|h_{j,k}|^2\}_{j\in\mathcal{J}}$, respectively. 
The inequality is from the fact that for $q\leq 1$ and $x\geq 0$, $e^{-qx}\leq \frac{q}{1+x}+1-q$, which    is tight for $x\ll 1$. 

Thus, the average probability of outage for each link is given by,
\begin{equation}\label{eq:poutapprox2}
p_{out}^k\leq 1-e^{-\frac{\gamma d_k^{\alpha}}{P_k}(N_0+J_k)},
\end{equation}
where $J_k=q\sum_{j\in \mathcal{J}} P_j/d_{j,k}^{\alpha}$. The cost of the minimum energy  path $\Pi$  in this case can be found by a similar derivation as in Section \ref{sec:optimalcost},
\begin{equation}\label{eq:pathcost}
\mathcal{C}(\Pi)=\frac{1}{\epsilon}\left( \sum_{\ell_k\in\Pi}\sqrt{d_k^{\alpha}(N_0+J_k)}\right)^2,
\end{equation}
where $\epsilon=\frac{-\ln(1-\pi)}{\gamma}$. 
Hence, by employing an estimate of the average jamming power obtained from recent channel measurements,   assigning the link cost $\mathcal{C}(\ell_k)=\sqrt{d_k^{\alpha}(N_0+J_k)}$ to each potential link $\ell_k$ in the network, and applying the routing algorithm discussed in the previous section, the efficient route can be found.
\section{Simulation Results}\label{sec:numerical}
We consider a wireless network in which  $n$ system nodes and $n_j$ jammers are placed uniformly at random on a $d\times d$ square. We assume that the closest system node to point $(0,0)$ is the source and the closest system node to the point $(d,d)$ is the destination.
 
 Our goal is to find an energy efficient route between the source and the destination. 
We assume that the threshold $\gamma=1$ (corresponding to throughput $\rho=1$), and the noise power $N_0=1$.
To analyze  the effect of propagation attenuation on the proposed algorithms, we consider  $\alpha=2$ for free space, and $\alpha=3$ and $\alpha=4$ for  terrestrial wireless environments.


{Because of the use of an approximation to obtain (\ref{eq:pout}), the  route obtained by MER-AP is not the absolute minimum energy route. 
However, in the following subsection we show that the gap between MER-AP and the exact (optimal) solution obtained by brute-force search 
is small.}

{{\subsection{Comparison with Optimal Algorithm}}}
{In this section, we perform an exhaustive search to obtain the optimal path. 
		Recalling the end-to-end outage probability given a path $\Pi=\left\langle \ell_1,\ldots,\ell_{|\Pi|}\right\rangle $, 
		\begin{align}\nonumber 
		p_{out}^{SD}(\Pi)=1-\prod_{\ell_k\in\Pi}\frac{e^{-\frac{\gamma N_0d_{k}^\alpha}{P_{k}}}}{\prod_{j\in \mathcal{J}}\left(1+\frac{\gamma P_{j}/d_{j,k}^\alpha}{P_{k}/d_{k}^\alpha}\right)}=\pi,
		\end{align}
		the optimization problem will be,
		\begin{align}
		\min \: \sum_{\ell_k\in\Pi} P_k
		\end{align}
		subject to,
		\begin{align*}
		 \sum_{\ell_k\in\Pi}\frac{\gamma N_0d_{k}^\alpha}{P_{k}}+\sum_{\ell_k\in\Pi}\sum_{j\in\mc{J}}	\log\left(1+\frac{\gamma P_{j}/d_{j,k}^\alpha}{P_{k}/d_{k}^\alpha}\right)=-\log(1-\pi),
		\end{align*}
		and,
		\begin{align*}
		P_k\geq 0,\: k=1,\ldots,|\Pi|.
		\end{align*}
		The constraint is convex and thus this problem has a local minimum. Hence, using any 
		nonlinear optimization program, we can obtain the minimum energy consumption of a given path. 
		In order to find the optimum path with minimum energy consumption, we should repeat this procedure for 
		any possible source-destination path in the network i.e.  $2^n$ times ($n=$number of relay nodes).
		}
		
\begin{figure}
	\begin{center}
		\includegraphics[width=.45\textwidth]{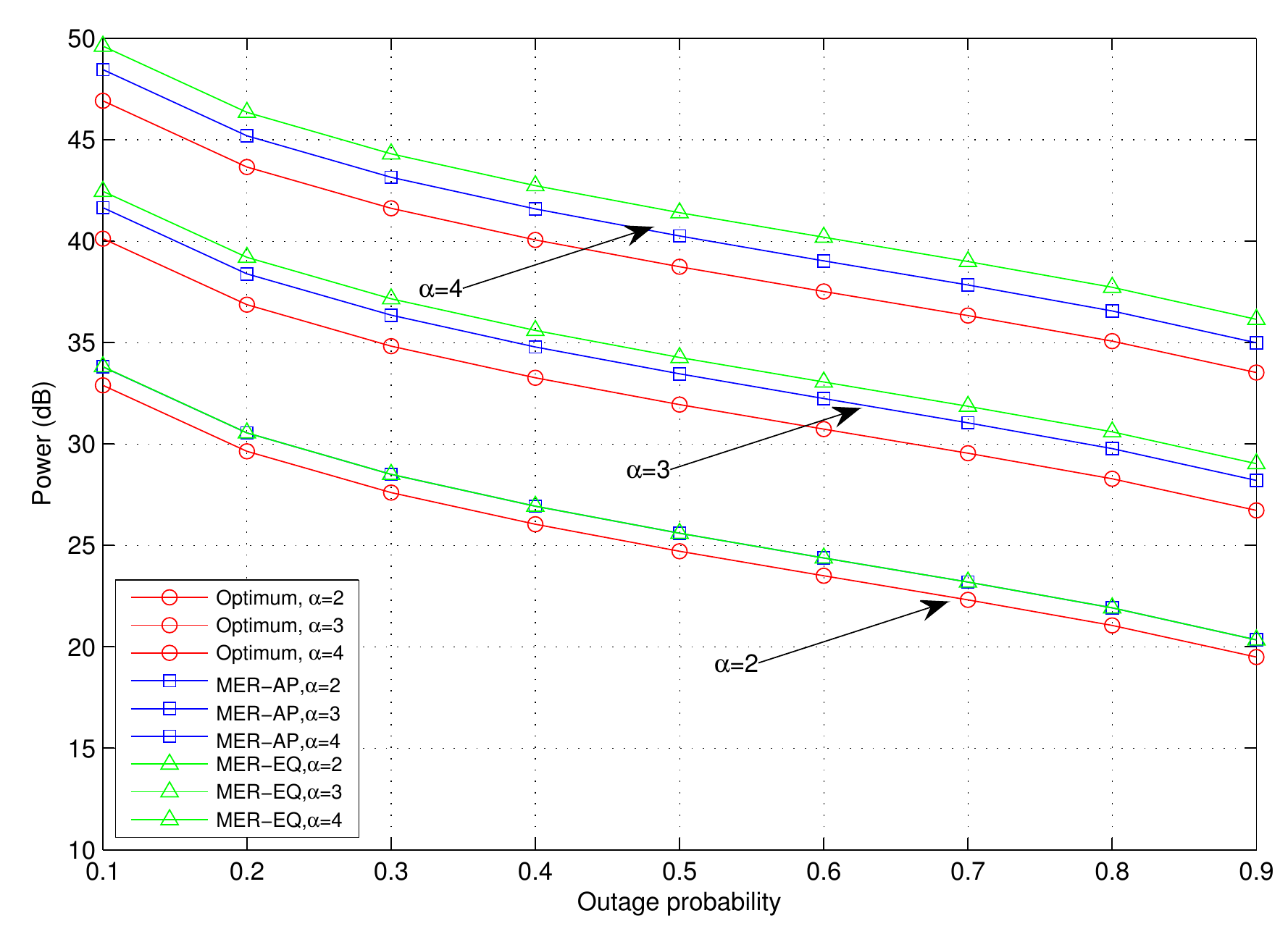}
	\end{center}
	\caption{{Totoal power versus outage probability for brute-force 'exact' algorithm (optimum), MER-AP, and MER-EQ.}}
	\label{fig:compare_OP_EQ_BF}
\end{figure}

	For a small network with  $n=8$ system nodes and $n_j=8$ jammers, Fig. \ref{fig:compare_OP_EQ_BF} shows the average energy spent using an exhaustive-search algorithm (optimum), and our proposed ``sub-optimal but efficient'' algorithms in the same network.
	 The results are presented in Fig. \ref{fig:compare_OP_EQ_BF}. 
	 As can be seen, the gap between the performance of the optimal solution and that of the proposed algorithm is small at less than $2$ dB. 
	 Moreover, we observe that MER-AP always outperforms MER-EQ
	that allocates equal outage probabilities to all links of a path.

In the rest of this section, we  show  that MER-AP finds a route that  takes  detours  to bypass the jammers  effectively and also allocates  suitable amounts of power to the transmitters  in such a way that it results in significant energy savings compared to MER-EQ.

\begin{figure}[h]
	\centering
	\includegraphics[width=.45\textwidth]{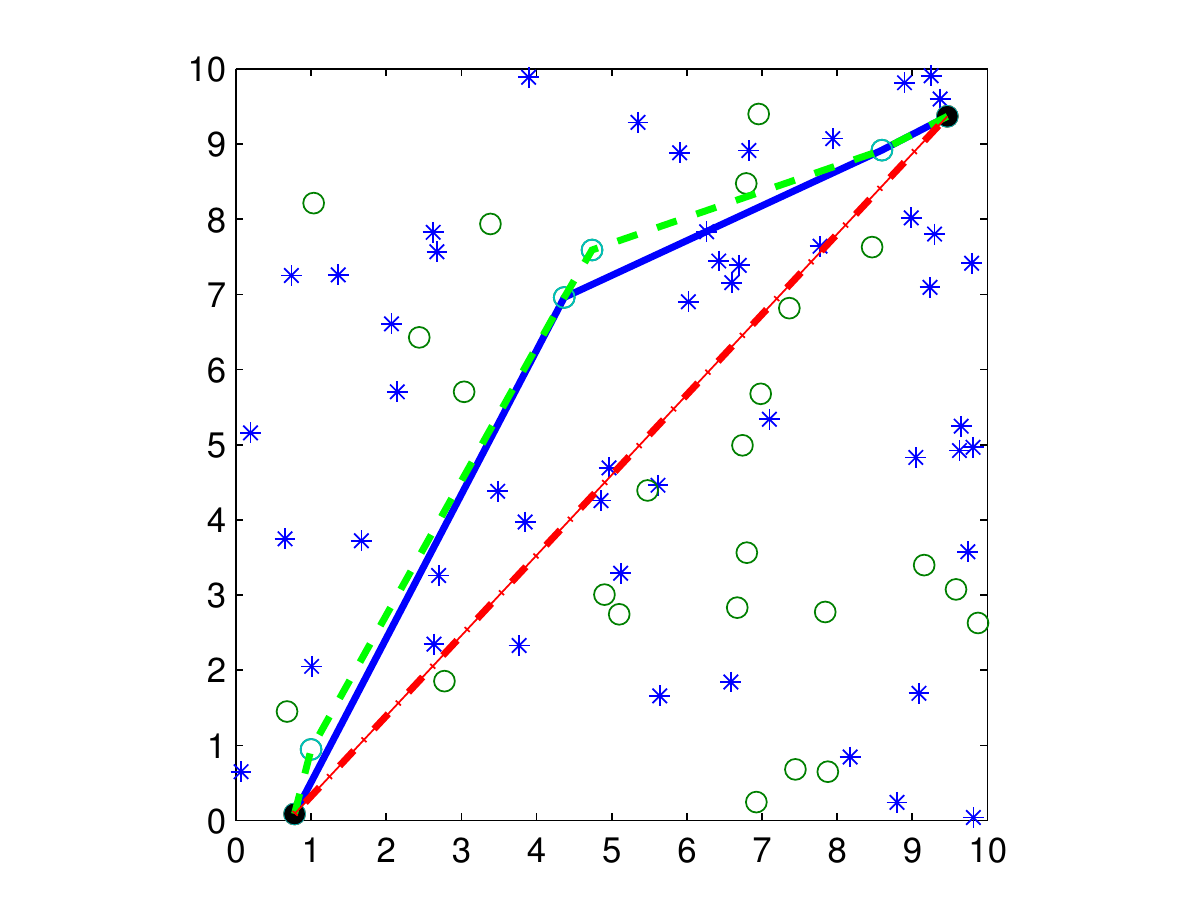}
	\caption{A snapshot of the network when $n=30$ system nodes (shown by circles) and $n_j=50$ jammers (shown by *) are placed uniformly at random. The transmit power of each jammer $P_j=1$, the target end-to-end outage probability $\pi=0.1$, and the path-loss exponent $\alpha=2$. The  MER-AP path is shown by the dashed line (green), the  MER-EQ path is shown by solid line (blue), and the MER route  is shown by the dash-dotted line (red). The  energy saved in this network for MER-AP   is 63.57\% and for MER-EQ is 54.47\% .}
	\label{fig:a2_random}
\end{figure} 

For the benchmark routing algorithm, we consider  a  minimum energy routing (MER) algorithm from the source to the destination with end-to-end target outage probability $\pi$.  
The MER algorithm is described in the following subsection.

\subsection{MER: Minimum Energy Routing} \label{sec:MER}
Consider
a wireless network with a source, a destination, and some other nodes that can be used as relays (without jammers).
The goal is to convey the message with minimum
aggregate power such that an end-to-end outage probability is guaranteed. 
 The outage probability of link $\ell_k$ is given by, 
\begin{equation}\label{eq:a1}
p_{out}^{k}= 1-\exp\left(\frac{-\gamma N_0 d_{k}^\alpha}{P_{k}}\right).
\end{equation}
Using the technique presented in Section IV, the optimal cost of path $\Pi$ is given by: 
\begin{equation}\nonumber
\mathcal{C}(\Pi)=\frac{1}{\epsilon}\left( \sum_{\ell_k\in\Pi}\sqrt{d_k^{\alpha}}\right)^2.
\end{equation}
Hence, we assign the link cost
$\mathcal{C}(\ell_k)=\sqrt{d_k^{\alpha}}$
to each potential link $\ell_k$ in the network and apply    Dijkstra's algorithm to find the optimum route.  

By using the
MER algorithm, the minimum energy route, the
outage probability of each link, and the transmit power of the source and each intermediate
relay on this route can be found. Now suppose an adversary spreads a number of jammers
in the network. In this case, we do not change the source-destination route and the outage
probabilities that are allocated to the links that belong to this route. However, because of
the interference due to the jammers at each receiver, the transmitters need to increase their
transmit power to have the same per link outage probability as when the jammers were
not present.
Since the channel gains between jammers and system nodes are  exponentially distributed, the average outage probability at each receiver of route $\Pi$ is given by (see the derivation presented in Section II for the link outage probability):
\begin{equation}
p_{out}^{k}=1-\frac{e^{-\frac{\gamma N_0d_{k}^\alpha}{P_{k}}}}{\prod_{j\in \mathcal{J}}\left(1+\frac{\gamma P_{j}/d_{j,k}^\alpha}{P_{k}/d_{k}^\alpha}\right)}.
\end{equation}
This equation can be solved numerically to find the required power of each link $\{P_k\}_{\ell_k\in\Pi}$ in the presence of  jammers.
As in the other approaches described earlier, the aggregate transmit power of the MER algorithm in the presence of jammers is considered as the cost of the scheme.
\subsection{Performance Metric}
Our performance metric is the \textit{energy saved} due to the use of  each algorithm. The  energy saved is defined as  the reduction in the  energy consumption of the system nodes when each algorithm is applied with respect to the  energy consumption when system nodes use the benchmark algorithm (i.e. MER). 

A snapshot of the network  when $n=30$, $n_j=50$, $P_j=1$, $\pi=0.1$, and $\alpha=2$ is shown in Fig. \ref{fig:a2_random}. The MER-AP path, MER-EQ path, and MER path are plotted in this figure. 
The percentage of energy saved in this example for  MER-AP is 63.57\% and for  MER-EQ  is 54.47\%.
As can be seen, using the MER-AP algorithm is more energy efficient than MER-EQ.

\begin{figure}
\begin{center}
 \includegraphics[width=.45\textwidth]{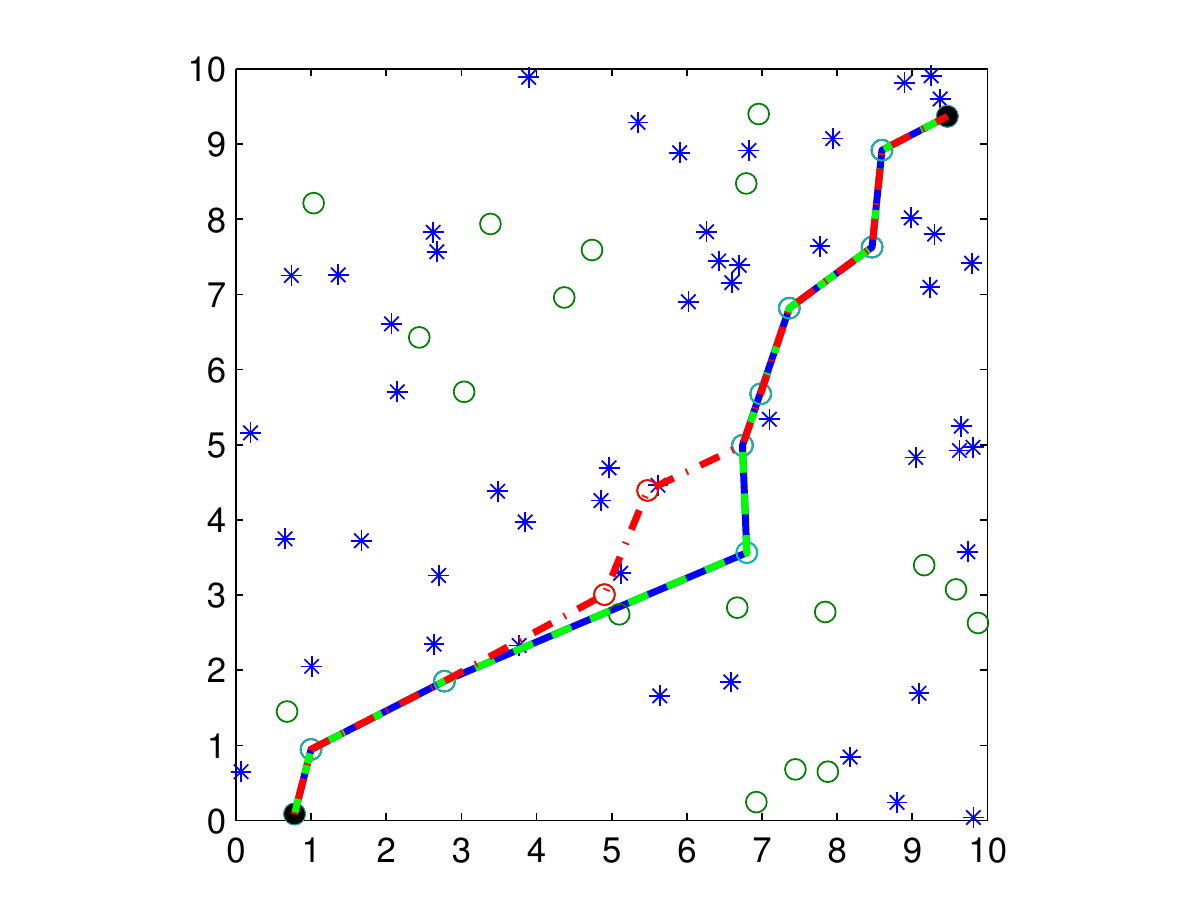}
 \end{center}
  \caption{A snapshot of the network with the same system node and jammer placement as in Fig. \ref{fig:a2_random}. Transmit power of each jammer $P_j=1$, target outage probability $\pi=0.1$, and transmission in a lossy environment is considered ($\alpha=4$). The  MER-AP path  is shown by the dashed line (green), the  MER-EQ path is shown by the solid line (blue), and the MER path is shown by the dash-dotted line (red). The  energy saved in this network for MER-AP  is 93.54\% and for MER-EQ  is 88.21\%.}
 \label{fig:a4_random}
 \end{figure} 

The MER-EQ, MER-AP, and MER paths for the same placement of the system nodes and jammers as in the   networks of Fig. \ref{fig:a2_random} for a higher path-loss exponent ($\alpha=4$) are shown in Fig. \ref{fig:a4_random}. In this case, the  energy saved  for MER-AP  is 93.54\% and for MER-EQ is 88.21\% . 
Note that although in this case the MER-AP algorithm and the MER-EQ algorithm both choose the same route, the percentage of energy saved using the latter approach is smaller, because we force all links in the path to have the same outage probability.
This shows the superiority of MER-AP algorithm over MER-EQ algorithm, as is also discussed in Section \ref{sec:disc}.

 In the sequel, we average our results over  randomly generated networks. The performance metric is the \textit{average energy saved}, where the averaging is over 100 random realizations of the network.
We consider the effect of various parameters of the network on the average energy saved by using the MER-AP and MER-EQ algorithms. 
 \subsection{Number of Jammers}
\begin{figure}[t]
\begin{center}
 \includegraphics[width=.45\textwidth]{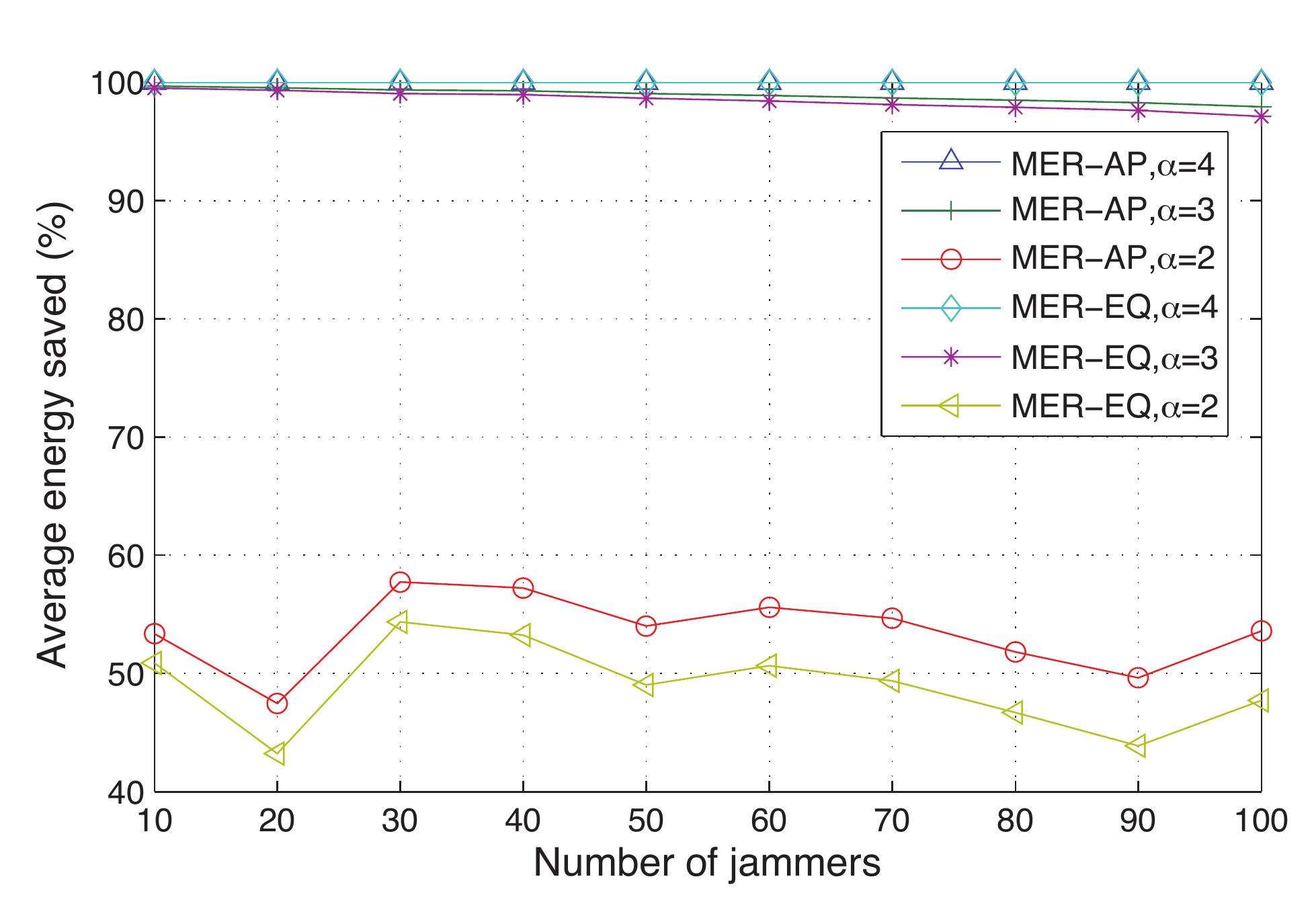}
 \end{center}
 \vspace{-.1 in}
 \caption{Average energy saved vs. number of static jammers for different values of the path-loss exponent. The transmit power of each jammer $P_j=1$, the end-to-end target probability of outage $\pi=0.1$, and $n=20$ system nodes are considered. The system nodes and the jammers are placed uniformly at random over a $10\times 10$ square. }
 \label{fig:r_vs_nj}
 \end{figure}

The effect of the number of  jammers on the average energy saved for different values of the path-loss exponent is shown in Fig. \ref{fig:r_vs_nj}.
It can be seen that the performance of MER-AP algorithm is always superior to the performance of MER-EQ algorithm, which is because of the  constraint on the outage probability of each hop of MER-EQ.
For both algorithms the average energy saved is not sensitive to  the number of jammers. The fluctuations in this figure are due to the random generation of the network.
On the other hand, the effect of the path-loss exponent on the average energy saved is dramatic. For terrestrial wireless environments ($\alpha=3$ and $\alpha=4$), the average energy saved by both algorithms is substantially higher than for free space wireless environments ($\alpha=2$).
The reason is that in the  environment with a higher path-loss exponent, the effect of the jamming signal is  local and thus  the jamming aware routes can take detours to avoid the jammers and obtain much higher energy efficiency.

\subsection{Jamming Power}
\begin{figure}
\begin{center}
 \includegraphics[width=.45\textwidth]{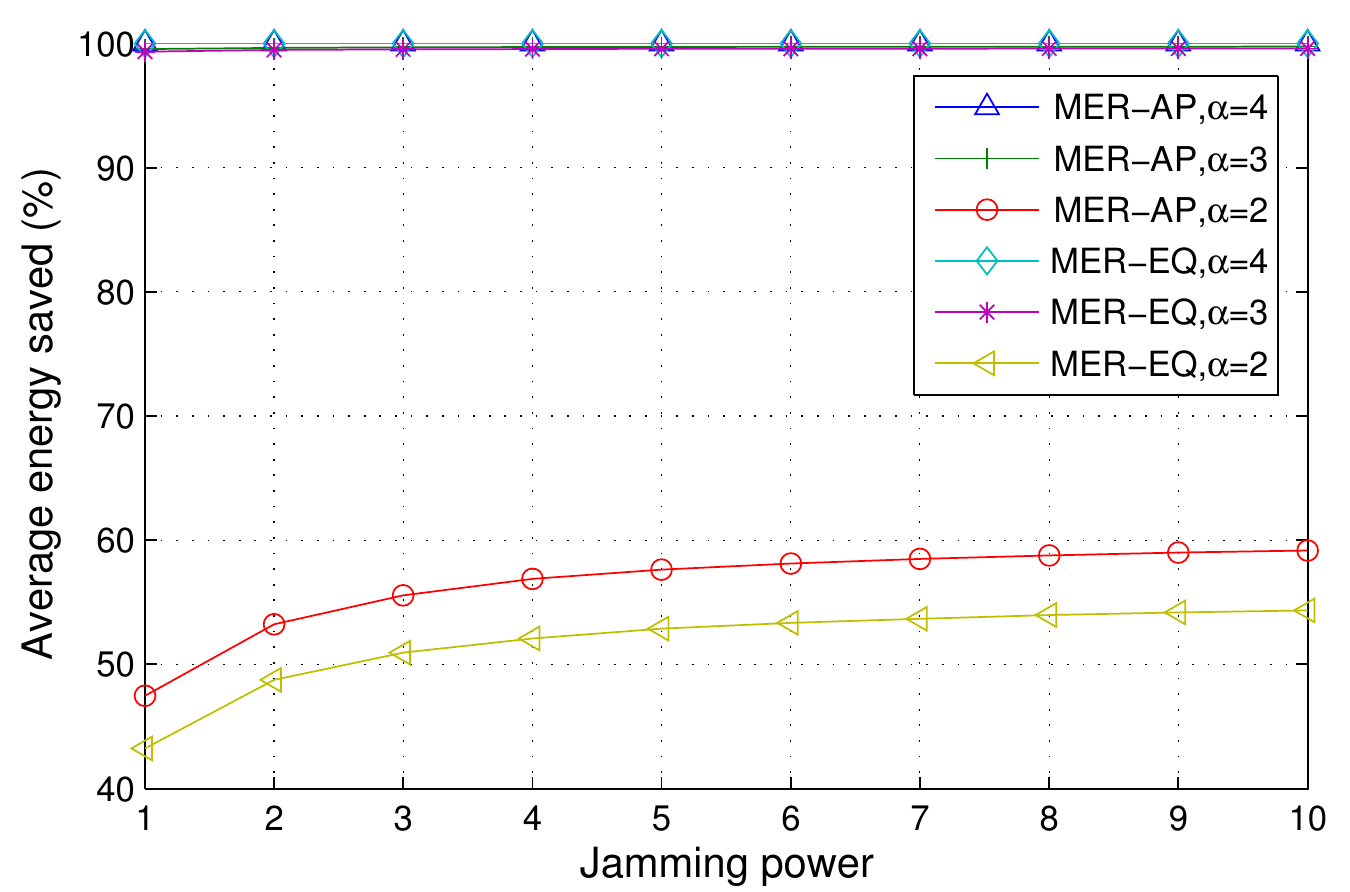}
 \end{center}
 \vspace{-.1 in}
 \caption{Average energy saved vs. jamming power of each malicious jammer for different values of the path-loss exponent.  $n_j=20$  number of jammers,  $n=20$ system nodes, and end-to-end target probability of outage $\pi=0.1$ are considered. The system nodes and the jammers are placed uniformly at random over a $10\times 10$ square. }
 \label{fig:r_vs_Pj}
 \end{figure}

The effect of jamming power on the average energy saved is shown in Fig. \ref{fig:r_vs_Pj}.
Again the energy efficiency of MER-AP algorithm is higher than that of MER-EQ algorithm due to better allocation of the per-link outage probabilities.
As the jamming power increases, the percentage of the energy saved by using both algorithms increases. 
Clearly, when the jamming power is higher,  the impact of jamming on  communication is greater, and thus  bypassing the jammers can lead to more energy efficiency of the routing algorithm.

\subsection{Size of Network}
\begin{figure}
\begin{center}
 \includegraphics[width=.45\textwidth]{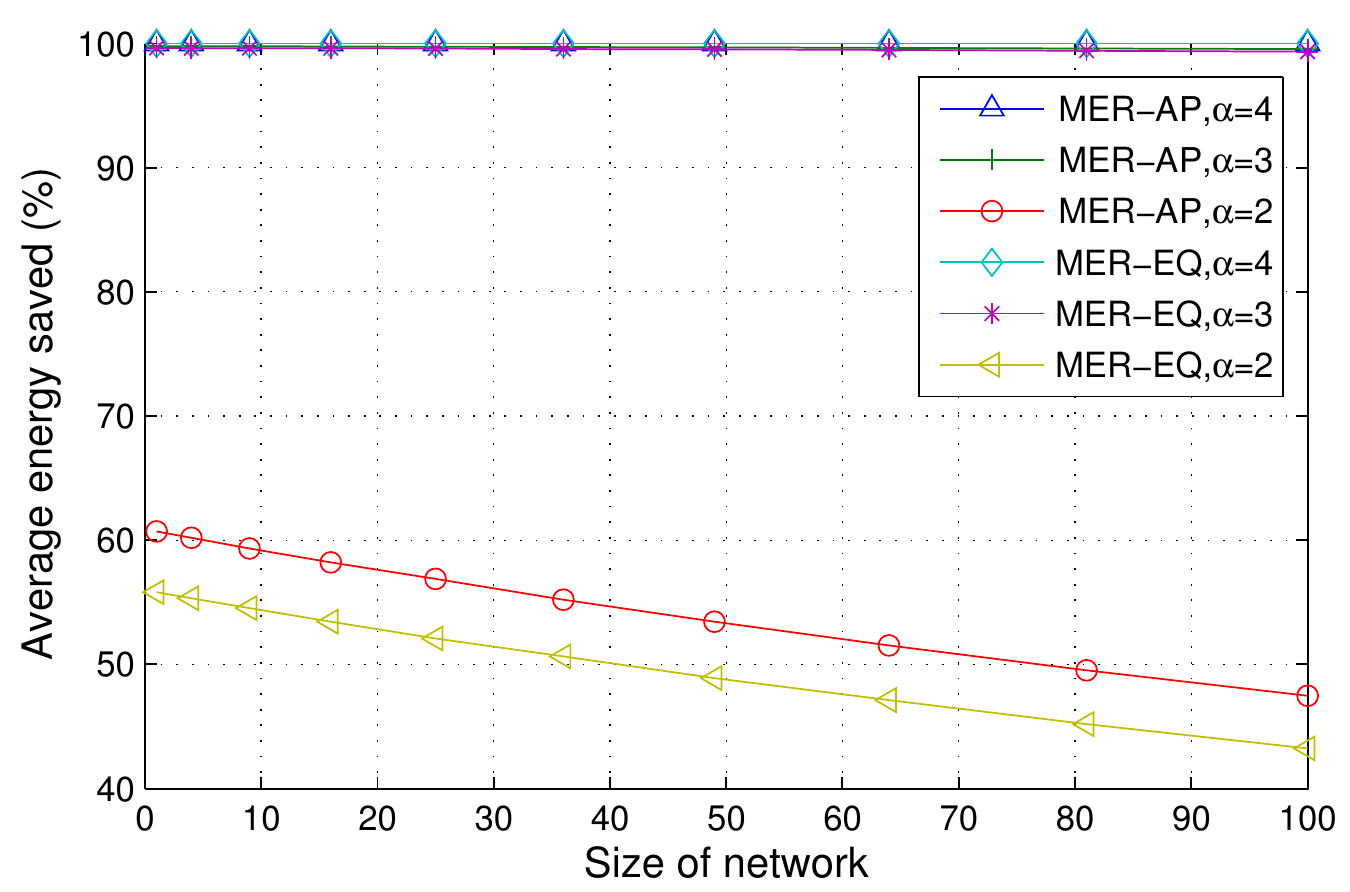}
 \end{center}
 \vspace{-.1 in}
 \caption{Average energy saved vs. area of the network for different values of the path-loss exponent.  The transmit power of each jammer $P_j=1$, $n_j=20$  number of jammers,  $n=20$ system nodes, and end-to-end target probability of outage $\pi=0.1$ are considered.  }
 \label{fig:r_vs_size}
 \end{figure}

The average energy saved versus the size of the network is shown in Fig. \ref{fig:r_vs_size}, where the area of the network changes from a $1\times 1$ square to a $10\times 10$ square. 
The average energy saved for terrestrial wireless environments for both algorithms is nearly $100\%$. 
When  free space parameters are used ($\alpha=2$), MER-AP algorithm always has a better performance than MER-EQ algorithm.
Also, it can be seen that  the percentage of the energy saved of using both algorithms is higher for smaller network areas.
The reason is that in a smaller network, the effect of jamming on the communication is higher and thus taking a route that bypasses the jammers helps more to improve the energy efficiency.

\subsection{Outage Probability}
\begin{figure}
\begin{center}
 \includegraphics[width=.45\textwidth]{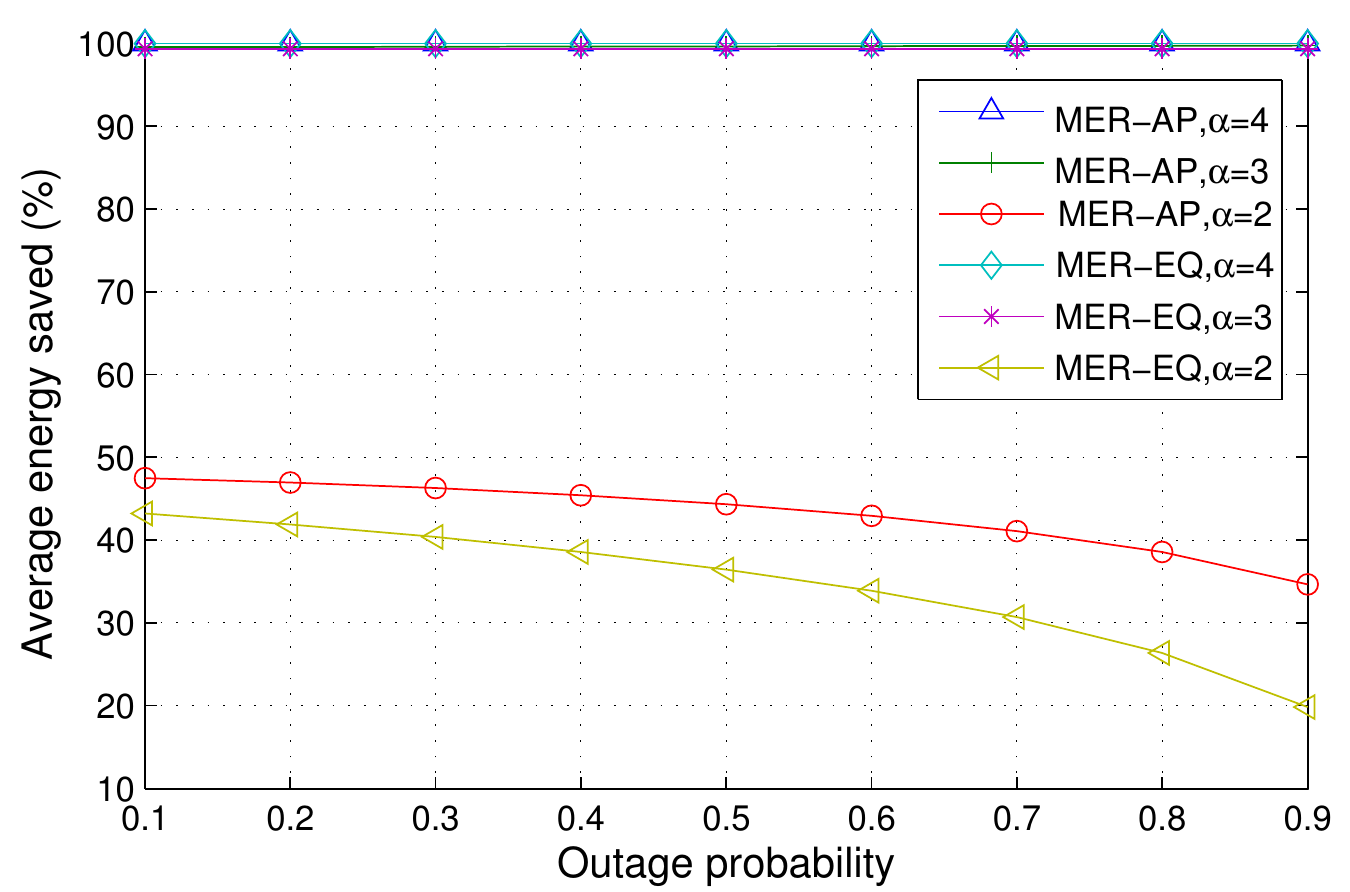}
 \end{center}
 \vspace{-.1 in}
 \caption{Average energy saved vs. end-to-end outage probability ($\pi$) for different values of the path-loss exponent. The transmit power of each jammer $P_j=1$, and $n_j=20$ jammers  and $n=20$ system nodes are considered. The system nodes and the jammers are placed uniformly at random over a $10\times 10$ square. }
 \label{fig:r_vs_pout}
 \end{figure}

In Fig. \ref{fig:r_vs_pout}, the percentage of average energy saved versus the outage probability is shown. For $\alpha=3$, and $\alpha=4$, the average energy saved is always very close to $100\%$. For $\alpha=2$, as the outage probability increases, more outages in the communication are acceptable,  and thus lower power is needed to mitigate the effect of a jammer close to the communication link. 
Hence, when the outage probability is greater, the percentage of energy saved by using a better path is less than when the outage probability is smaller.

\subsection{Power Histogram}
\begin{figure}
\begin{center}
 \includegraphics[width=.45\textwidth]{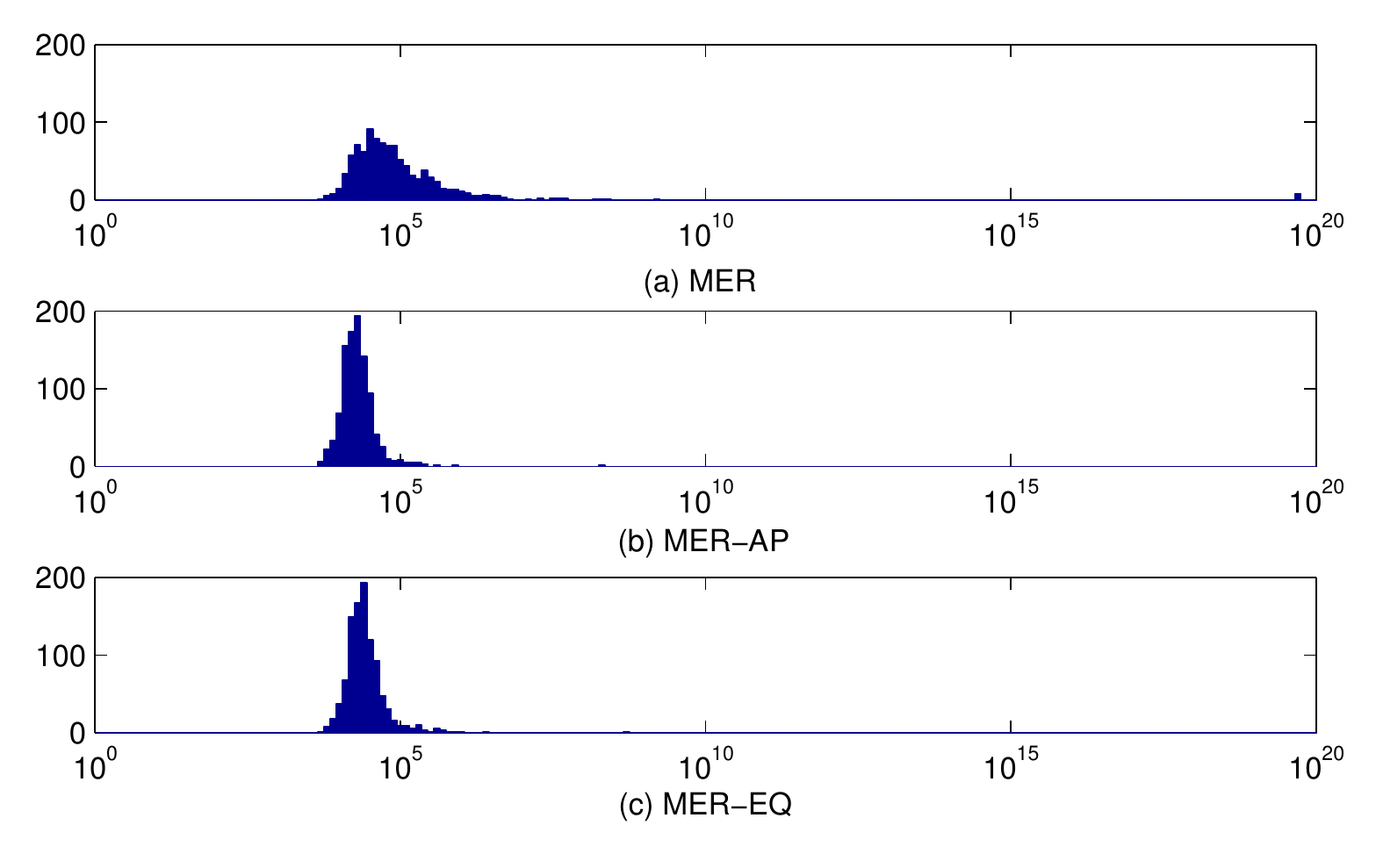}
 \end{center}
 \caption{The histograms of the number of network realizations versus  cost of transmission (aggregate power)  for (a) MER, (b) MER-AP, and (c) MER-EQ  are shown. The system nodes and the jammers are placed uniformly at random over a $10\times 10$ square,  where $\alpha=3$, $\pi=0.1$, $n=20$, and $n_j=50$.
 For the benchmark,  the values of the total cost are scattered, and the average
energy is dominated by a few bad realizations, while for (b) and (c), the values of
the total cost are concentrated around a central value (here
$10^{4}$). }
\label{fig:histo}
 \end{figure}

To further investigate the enormous gains in average energy for
higher values of $\alpha$, the histograms of the number of network realizations
versus the total cost of transmission (aggregate power) for
(a) MER algorithm, (b) MER-AP  algorithm, and (c) MER-EQ algorithm for $10^{3}$ realizations
of the network are shown in Fig. \ref{fig:histo}. In this figure $\alpha=4$,
$\pi=0.1$, $n=20$, and $n_j=30$.  For the MER, it can be seen
that the values of the total cost are scattered, and the average
energy is dominated by a few bad realizations.  On the
other hand, when MER-AP and MER-EQ are used, the values of
the total cost are concentrated around a central value (here
$10^{4}$). This explains the large gains in average energy shown in
previous sections, and also indicates that the MER-AP and MER-EQ are robust
against changes in the system node and jammer placements.
\subsection{Network Throughput}
When  MER-AP is used, we expect  the network can achieve a higher  throughput, since the transmit powers of the nodes in the efficient path are smaller, and thus more nodes can transmit their messages simultaneously.  
To study network throughput, in this section, we simulate multiple concurrent flows in the network and implement scheduling in addition to routing.
The maximum  throughput for a given number of concurrent flows can be obtained as follows.

\textit{Scheduling problem.}
Consider a subset $S \subseteq \mc{L}$ of the links. 
We call $S$ a ``transmission set'' if all links in $S$ can be scheduled concurrently. 
Moreover, $S$ is a ``maximal'' transmission set if it cannot be grown further. Let $\mc{S} = \set{S_1, \ldots, S_M}$ denote the set of all maximal transmission sets of the network. A schedule is specified by a set of weights ${\boldsymbol\alpha}=\set{\alpha_1, \ldots, \alpha_M}$, where each weight $0 \le \alpha_i \le 1$ specifies the fraction of time for which the maximal transmission set $S_i$ is scheduled\footnote{We assume a time slotted system where each time slot is of unit length. The weights $\alpha_i$ specify the fraction of time each set $S_i$ is scheduled in a time slot using a TDMA scheduler.}. It follows that $\sum_{i=1}^M \alpha_i = 1$ for a feasible schedule. In general, there is an exponential number of maximal transmission sets in a network and finding them is an NP-hard problem \cite{lawler1980generating}.

\textit{Maximal transmission sets.}
To obtain a practical approximation, we can use only a subset of all maximal transmission sets. As we increase the number of maximal transmission sets, the accuracy of the approximation increases. Algorithm~\ref{a:optimal} is used repeatedly to obtain a subset of all maximal transmission sets. 

\begin{algorithm}
\caption{Maximal Transmission Sets}
\begin{algorithmic}[1]
\STATE $S \leftarrow \set{}$
\WHILE{$\mc{L} \neq \set{}$}
	\STATE Choose $\ell_i \in \mc{L}$ at random
	\STATE $\mc{L} \leftarrow \mc{L} \backslash\set{\ell_i}$
	\IF{$\ell_i$ is schedulable with $S$}
		\STATE $S \leftarrow S \cup \set{\ell_i}$
	\ENDIF
\ENDWHILE
\RETURN $S$
\end{algorithmic}
\label{a:optimal}
\end{algorithm}

\textit{Throughput.}
Suppose there are $L$ flows in the network denoted by $\mc{F}=\set{f_1, \ldots, f_L}$. Let $x_i$ denote the rate of flow $f_i$ and $\mc{X}=\set{x_1, \ldots, x_L}$. The  path computed for flow $f_i$ is denoted by $\Pi_i$. Our goal is to compute the maximum flow rate in the network. Let $\lambda$ denote the capacity of link $\ell_k$, which is a constant for every link in the network (this is ensured by our power allocation algorithm).
\begin{itemize}
\item
The total flow rate that passes through link $\ell_k$ is given by,
\[
	\sum_{\forall f_i \in \mc{F}:\, \ell_k \in \Pi_i} x_i
\]
\item
The total capacity of link $\ell_k$ adjusted for scheduling is given by,
\[
	\lambda \cdot \sum_{\forall S_i \in\mc{S}:\, \ell_k \in S_i} \alpha_i
\]
\end{itemize}

\begin{figure}
\begin{center}
 \includegraphics[width=.45\textwidth]{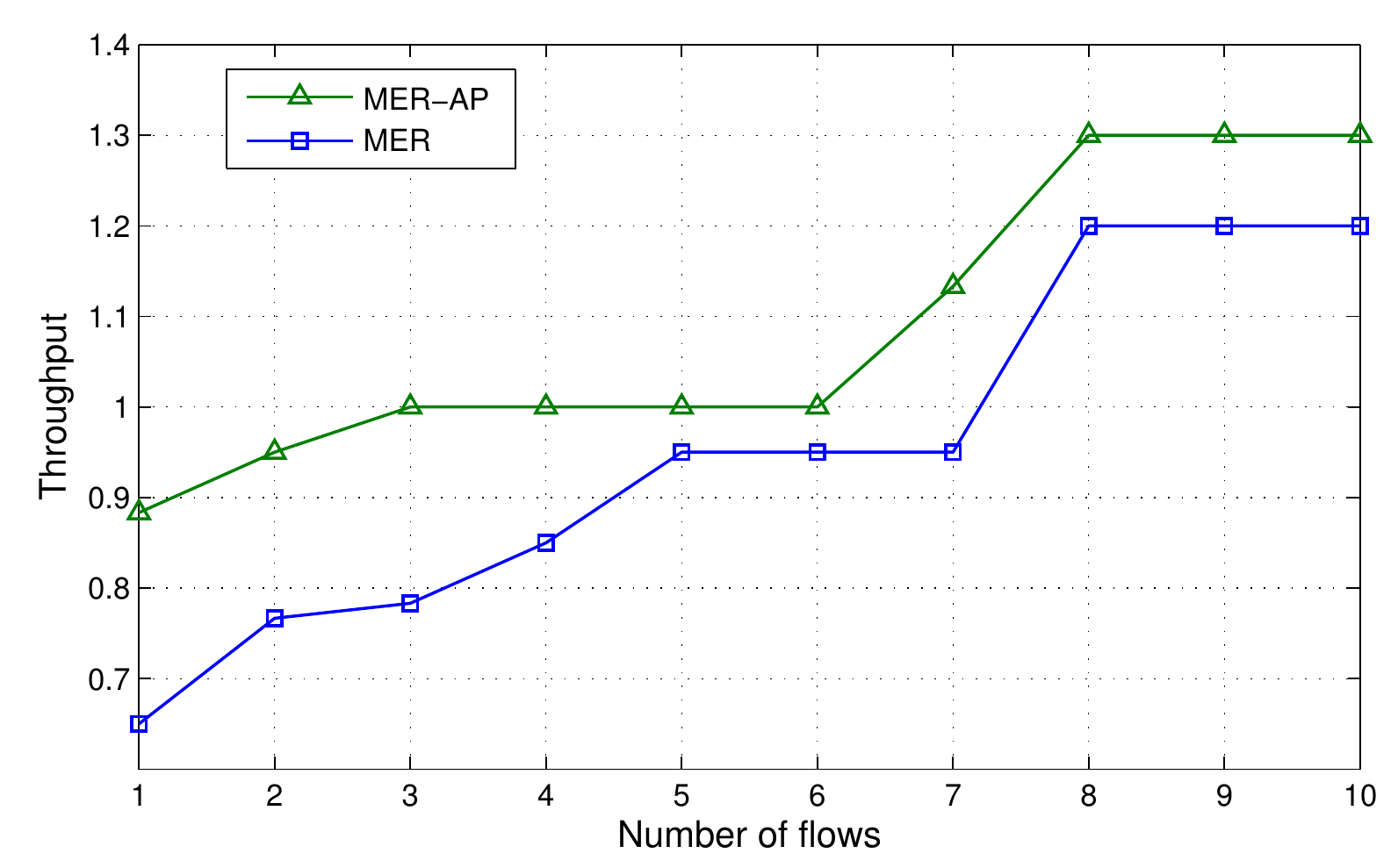}
 \end{center}
 \caption{Throughput versus the number of concurrent flows for the MER-AP algorithm and for MER, when
 the end-to-end outage probability is $\pi=0.2$, and $n=10$ system nodes and $n_j=20$ jammers are present. 
 }
\label{fig:rate_vs_nflow}
 \end{figure}

To compute the maximum throughput, one has to solve the following optimization problem:
\begin{align}
	& \max \sum_{f_i \in \mc{F}} x_i\\
	& \text{subject to:}\nonumber\\
	& \sum_{\forall f_i \in \mc{F}:\, \ell_k \in \Pi_i} x_i \le
		\lambda \cdot \sum_{\forall S_i \in\mc{S}:\, \ell_k \in S_i} \alpha_i\\
	& \qquad \sum_{\alpha_i \in {\boldsymbol \alpha}} \alpha_i = 1\\
	& \qquad \alpha_i \ge 0
\end{align}
Since the constraints as well as the objective function are linear, the above problem is a convex optimization problem if the routes $\Pi_i$ and maximal transmission sets $S_i$ are known.
We used Matlab to solve this optimization problem and compute the total throughput.
The throughputs versus the number of concurrent flows for MER-AP and for MER are shown in Fig. \ref{fig:rate_vs_nflow}.
 The end-to-end outage probability is $\pi=0.2$, where $n=10$ system nodes and $n_j=20$ jammers are present. 
As expected,  the MER-AP can achieve higher throughput than the MER algorithm. 

\begin{figure}
\begin{center}
 \includegraphics[width=.45\textwidth]{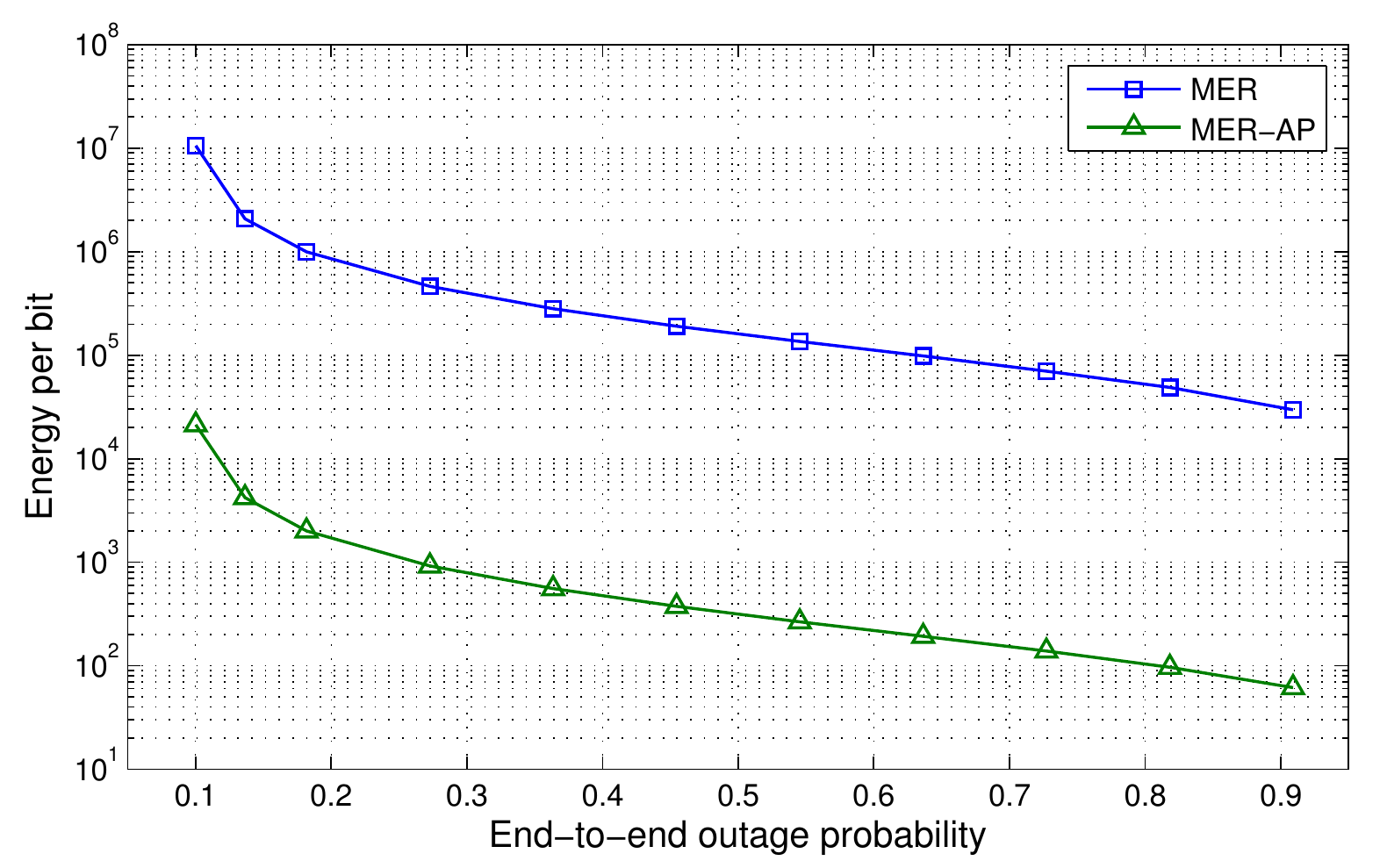}
 \end{center}
 \caption{Energy per bit versus the end-to-end outage probability for MER-AP and MER.  The  throughput is measured for five concurrent flows, where $n=10$ system nodes and $n_j=20$ jammers are present in the system. The amount of energy per bit MER-AP uses is about two orders of magnitude less than MER. }
\label{fig:energy_per_bit}
 \end{figure}

\textit{Energy per bit.} In order to compare the amount of energy each algorithm needs to obtain the throughput shown in Fig. \ref{fig:rate_vs_nflow}, the energies per bit versus the outage probability for MER-AP and MER are shown in Fig. \ref{fig:energy_per_bit}. Energy per bit  is obtained by dividing the total power consumed by the system nodes divided by the maximum throughput of the network for a given number of flows. In this figure, the maximum throughput when five concurrent flows exists in the network, where $n=10$ system nodes and $n_j=20$ jammers are present, is plotted. 
As expected,  in both algorithms for higher outage probabilities less energy per bit is required. 
Also, the amount of energy per bit MER-AP uses is about two orders of magnitude less than the amount of energy per bit MER consumes.
 \subsection{Dynamic Jammers}
\begin{figure}
\begin{center}
	\includegraphics[width=.44\textwidth]{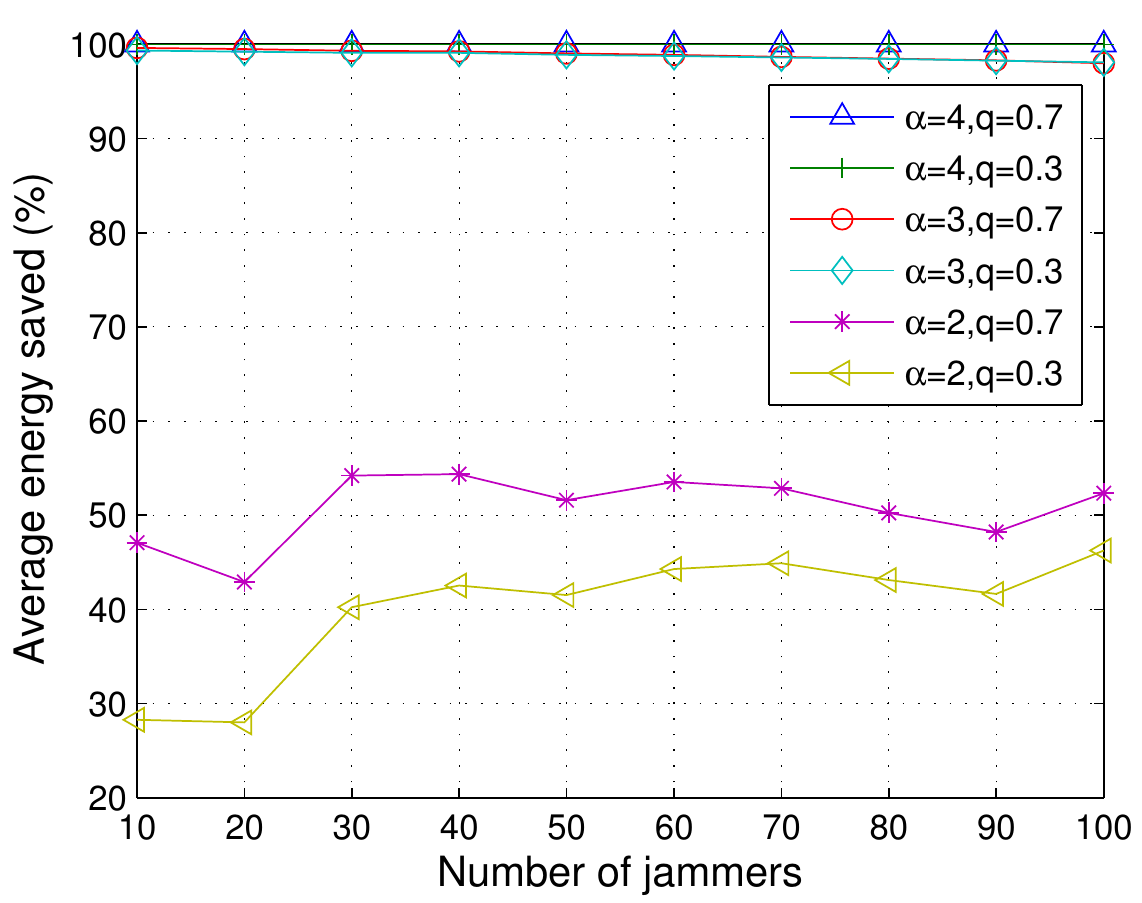}
 \end{center}
 \caption{Average energy saved vs. number of dynamic jammers for different values of the path-loss exponent and probability $q$ of a jammer being in ``ON'' state. The transmit power of each jammer $P_j=1$, the target end-to-end outage probability $\pi=0.1$, and $n=20$ system nodes  are considered. The system nodes and the jammers are placed uniformly at random over a $10\times 10$ square. }
 \label{fig:r_vs_nj_dynamic}
 \end{figure}

In this section, we investigate  the effect of the number of dynamic jammers  on the average energy saved when employing MER-AP.
 The average energy saved versus number of jammers for probability of a jammer being  ``ON'', $q=0.3$ and $q=0.7$, and for various values of the path-loss exponent, $\alpha=2, 3,4$, are considered in Fig. \ref{fig:r_vs_nj_dynamic}. 
The simulations are done over $100$ random realizations of the network.
As can be seen, the average energy saved is again insensitive to the number of jammers (the fluctuations in this figure are due to the randomness of the network realizations). 
For $\alpha=2$,  the percentage of energy saved is higher  when $q$ is greater, since the effect of jammers on the network is greater and thus, by using MER-AP algorithm and bypassing the jammers, a higher energy efficiency can be gained. 
 For terrestrial wireless environments, i.e. for $\alpha=3$ and $\alpha=4$, the average energy saved by using MER-AP is always substantial and close to $100\%$.

 
\section{Discussion}\label{sec:disc}

\subsection{Distributed Implementation}
It is useful to mention that distributed implementation of the algorithms presented in this paper is straightforward.
The link costs introduced in previous sections can be calculated locally by using the average of the total jamming signal at each node,  and this information can be passed to  neighboring nodes.
Then, any distributed distance vector routing technique such as the Bellman-Ford algorithm can be used  to find the minimum energy path. 
{\subsection{Retransmission-aware Routing}}
{Many of energy-aware routing algorithms  use  the energy spent in a
``single transmission" as their link cost metric, and do not consider retransmissions   \cite{tekbiyik2011energy}. There are many reasons for this. For instance, there are wireless  protocols that do not provide link layer retransmissions (e.g. CSMA, MACA \cite{zhu2006accurate}), and   there exist many situations where link-layer retransmissions are  harmful and degrade the overall quality of transmission (e.g. \cite{chan2008impact}).}
	
	{Nevertheless, Banerjee and Misra argued that a single-transmission
	formulation does not capture the total energy
	spent to reliably deliver a message, and thus the cost should include the energy spent in retransmissions as well \cite{banerjee2002minimum}.
	The special class of algorithms that consider the energy spent in retransmissions  are called ``retransmission-aware algorithms".
	The retransmission-aware algorithms are classified into two categories \cite{banerjee2002minimum}:
	\begin{enumerate}
		\item 
		End-to-End Retransmissions (EER model): relays do not perform link-layer retransmissions, and 
		retransmissions initiated by the source node ensure reliable message delivery.
		\item
		Hop-by-Hop Retransmissions (HHR model): each relay performs retransmissions to provide reliable
		forwarding to the next hop.
	\end{enumerate}
	When we have per-hop error probabilities, it is appropriate to consider the HHR model, because the number of retransmissions on a link is independent of the other links on the path,  hence making it tractable. 
	In our paper, we can use the HHR model with our 
	equal-outage-per-link (MER-EQ) algorithm since it considers per-hop outage probabilities.
	However, please note that  in all practical systems including 
	WiFi, there is a limit on the number of retransmissions to avoid long 
	delays. This means that all real systems require some bound on the 
	outage probability to successfully deliver packets. 
	If we consider a limit on the number of retransmissions, then with our 
	equal-outage algorithm, the outage per link is given by ${p_{out}}^N$ 
	where $N$ is the limit on the number of retransmissions and $p_{out}$ is the 
	equal target outage on each link.}
	
	{On the other hand, in \cite{banerjee2002minimum,banerjee2004energy} it is shown that  when we have per-hop probabilities, using the EER model makes the formulation analytically intractable for  minimum-cost path computation, and thus the minimum-energy path considering retransmissions cannot be found efficiently (i.e., in polynomial time).}
	
	{In  the MER-AP algorithm, since we consider end-to-end error probability and the weight of each link in the routing algorithm does not depend on the end-to-end outage probability, we can use the EER model and  the problem is 	tractable. 
	But, since we do not have the per-hop outage probabilities  in the formulation of MER-AP, it is intractable  to use this algorithm with the HHR model. 
	For the EER model, the total transmission energy can be expressed as \cite{banerjee2002minimum},
	\begin{equation}
	P_{total}=\frac{\sum_{i=1}^KP_i}{\prod_{i=1}^K\left( 1-p_{out}^{i}\right) }
	\end{equation}
	where $p_{out}^{i}$ is the outage probability of the $i^{th}$ link. We have,
	\[p_{out}^{SD}=1-\prod_{i=1}^K\left( 1-p_{out}^{i}\right) \] 
	Hence,
	\begin{equation}
	P_{total}=\frac{\sum_{i=1}^KP_i}{1-p_{out}^{SD}}
	\end{equation}
	For a fixed and-to-end outage probability, minimizing $P_{total}$ is equivalent to minimizing $\sum_{i=1}^KP_i$. Hence,
	in order to include the power spent on retransmissions, we should multiply the cost of a path by a factor $\frac{1}{1-p_{out}^{SD}}$.
	}
\section{Related Work}
\textbf{Spread Spectrum and Beamforming.}
Traditional methods to combat jamming attacks include spread spectrum and beamforming \cite{proakis2000digital,peterson1995introduction,scholtz1982origins,liu2010randomized,zhou2002digital}; however, these approaches are only a partial solution in the case of broadband jammers, jammers with directional antennas, or multiple jammers, and, as discussed in the Introduction,  these methods often simply  increase the cost of jamming.  
Nevertheless, our routing algorithms can be used in conjunction with these techniques to increase the robustness of the system against  jamming attacks.

\textbf{Other Jamming Evasion Techniques.} 
When the system nodes are able to move, they can simply leave the jammed area to a safe place. This is the basis of the spatial retreat technique, in which the system nodes move away from a stationary jammer \cite{xu2004channel,xu2006jamming}.
Another jamming evasion technique is channel surfing, where  the system nodes basically change their communication frequency to an interference-free frequency band when necessary \cite{xu2007channel}. 
These approaches, however, are orthogonal to the problem considered here which deals with static nodes.

\textbf{One-Hop Communication in the Presence of  Jamming.}
Several works consider one-hop energy aware communication  in the presence of one jammer \cite{mal2000analysis,li2007optimal,zhu2010stochastic,khan2011adaptive}. It is usually treated as a game between a jammer and two system nodes. The objective of the jammer is to increase the cost (energy) of  communication for the system nodes, whereas the objective of the system nodes is  increasing the cost of jamming for the jammer and  conveying their message with a minimum use of energy.
Unlike these approaches, in this work  we consider multi-hop communication in the presence of many jammers.

{\textbf{Routing  in the Presence of Jamming.}
Some works  consider jamming-aware  multi-path routing \cite{kushman2007r,mueller2004multipath,nasipuri1999demand,yang2006source,mustafa2012jamming}.   While in a completely different setting from this work, these multi-path algorithms are mostly based on
sending a message along multiple node-disjoint or link-disjoint
paths to ensure fault-tolerant message delivery. Although such algorithms are suitable for wired
networks, their application in wireless networks is challenging
due to lack of path diversity at the source or destination of
a communication session. In particular, in
wireless networks node-disjoint and link-disjoint paths are not
necessarily independent paths. Moreover, network topology, in wireless networks, is a function of power allocation at the physical-layer and propagation environment,
e.g., fading.}

\textbf{Energy-Aware Routing.}
In order to minimize energy consumption in wireless networks, numerous energy-efficient routing  algorithms have been studied \cite{singh1998power,rodoplu1999minimum,chang2000energy,kwon2006energy,dehghan2011minimum,manohar2003power}. {For instance, in \cite{manohar2003power}  energy-efficient routing with an end-to-end probability of error constraint is considered. However, \cite{manohar2003power} does not consider any kind of jamming
and/or spatially non-uniform interference.} Instead of   the total energy usage of the network nodes, some works consider the battery usage of each node, or balanced energy dissipation in the  network as their criteria \cite{shah2002energy,ganesan2001highly,chang2004maximum}. 
For example, in \cite{shah2002energy}, instead of choosing one source-destination path, the algorithm chooses several paths  and uses them alternatively to avoid quick energy depletion of each path.
While minimum energy routing has been studied extensively, a few works (e.g. see \cite{ghaderi2013efficient,ghaderi2014min})  considered security-aware  routing. However, unlike our work, they considered routing in the presence of passive eavesdroppers, which is different from the problem considered in this work with active jammers.

\section{Conclusions and Future Work}\label{sec:conc}

In this paper, we  considered minimum energy routing in a quasi-static multi-path fading environment and in the presence of multiple static and dynamic malicious jammers.  
The outage probability equation considering the jammers is intricate; thus, we established an approximation for the outage probability, based on which 
we developed an algorithm  to obtain a minimum energy path between a single source and a single destination with an end-to-end outage probability constraint. The algorithm requires only  the knowledge of the total average power received from the jammers at each system node over a long time period. 

By performing simulations using various network parameters, we compared the energy cost of  our algorithms to that of
a   jamming oblivious minimum energy routing algorithm, and showed  that our algorithms achieve significantly better energy efficiency.
In particular, it is shown that the energy saved by using our algorithms compared to the jamming oblivious scheme, especially in the case of terrestrial wireless networks with path-loss exponent $\alpha>2$, is  substantial.
The consideration of more sophisticated dynamic jammers with or without eavesdropping capabilities is a topic for further research.

\bibliographystyle{ieeetr}
\bibliography{mycite}

\end{document}